\documentclass{WileyMSP-template}
\usepackage{braket}
\usepackage{hyperref}
\usepackage{array}
\usepackage{booktabs}
\usepackage{subcaption}
\usepackage{amsfonts}
\usepackage{amsmath} 
\usepackage{ragged2e}
\usepackage{xcolor}

\begin{document}
\rhead{\includegraphics[width=2.5cm]{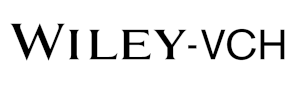}}

\title{Quantum Circuit Design using a Progressive \\Widening Enhanced Monte Carlo Tree Search\\}
\maketitle
\author{Vincenzo Lipardi*, Domenica Dibenedetto, Georgios Stamoulis, and \\Mark H.M. Winands}


\begin{affiliations}
Department of Advanced Computing Sciences, Maastricht University, Maastricht, \\The Netherlands\\
Email: vincenzo.lipardi@maastrichtuniversity.nl
\end{affiliations}


\begin{abstract}
The performance of Variational Quantum Algorithms (VQAs) strongly depends on the choice of the parameterized quantum circuit to optimize. One of the biggest challenges in VQAs is designing quantum circuits tailored to the particular problem. 
This article proposes a gradient-free Monte Carlo Tree Search (MCTS) technique to automate the process of quantum circuit design. Our proposed technique introduces a novel formulation of the action space based on a sampling scheme and a progressive widening technique to explore the space dynamically. When testing our MCTS approach on the domain of random quantum circuits, MCTS approximates unstructured circuits under different values of stabilizer R\'enyi entropy. It turns out that MCTS manages to approximate the benchmark quantum states independently from their degree of nonstabilizerness. Next, our technique exhibits robustness across various application domains, including quantum chemistry and systems of linear equations. Compared to previous MCTS research, our technique reduces the number of quantum circuit evaluations by a factor of 10 up to 100 while achieving equal or better results. In addition, the resulting quantum circuits exhibit up to three times fewer CNOT gates, which is important for implementation on noisy quantum hardware. 

\end{abstract}


\justifying
\section{Introduction}
The design of quantum algorithms to address practical problems in near-term quantum devices is a growing research field \cite{preskill, bharti2022noisy, dalzell2023quantum}. Variational Quantum Algorithms (VQAs) are among the most investigated methods for this purpose \cite{cerezo2021variational}. They are hybrid quantum-classical algorithms where the problem of interest is encoded into an optimization task over parameterized quantum circuits (PQCs). The computation is distributed between a quantum and a classical device. The evaluation of the objective function defined for quantum circuits is performed on the former, while the parameter optimization for the PQC is performed on the latter.
The extensive research on VQAs can be attributed to two main reasons. First, VQAs are a hardware-tailored approach as users can design them considering the specifics of the quantum hardware at their disposal, in terms of the native universal gate set, quantum circuit depth, and connectivity between qubits. This property is crucial given the current limitations of quantum devices. Second, VQAs provide a framework for dealing with a wide range of applications, including quantum chemistry and condensed matter physics \cite{kandala2017hardware}, combinatorial optimization \cite{farhi2014quantum}, quantum machine learning \cite{schuld2021machine}, linear algebra \cite{bravo2023variational,xu2021variational}, and finance \cite{stamatopoulos2020option}.

The choice of the PQC to optimize \cite{cerezo2021variational}, commonly known as \textit{ansatz}, is one of the biggest challenges in VQAs. Performance significantly depends on the ansatz chosen, which needs to be selected considering both the constraints imposed by the hardware and the structure of the problem. The goal is to design a PQC that is expressive and presents an easy-to-explore landscape for the parameter optimization task \cite{mcclean2018barren}.

\noindent In general, the design of quantum ansatz is a difficult task. However, it is possible to define guidelines to design `problem-inspired' ansatzes when the problem under study presents particular symmetries. This is the case in quantum chemistry, where different physics-inspired solutions have been proposed \cite{kandala2017hardware, lee2018generalized}. Nevertheless, there is no systematic approach for designing effective ansatz on general problems. The task may be addressed by brute-force classical algorithms, which scale exponentially with respect to the quantum circuit size \cite{nielsen2010quantum}. For this reason, Quantum Architecture Search (QAS), also known as \textit{quantum ansatz search} or \textit{quantum circuit design}, emerges as a crucial field to automatically explore the architectures of PQCs by leveraging computational resources \cite{grimsley2019adaptive, tang2021qubit, zhu2022adaptive}. 


This article presents the Progressive Widening enhanced Monte Carlo Tree Search (PWMCTS) developed to tackle the QAS problem. In contrast to previous MCTS works \cite{meng2021quantum, wang} that rely on strong assumptions on the problem structure,  PWMCTS does not require any prior knowledge of the problem. Moreover, it employs a gradient-based technique only on the final PQC to fine-tune the parameters but not iteratively along the tree search as in previous MCTS works. Furthermore, we provide a different tree formulation of the QAS problem and employ a different variant of MCTS for the search.

\noindent The three main contributions of this article are as follows.
\begin{enumerate}
    \item We develop a gradient-free approach to QAS based on MCTS. It consists of a sampling technique defined for PQCs that structures the action space as an infinite and discrete space \cite{franken2022quantum} explored using a progressive widening technique \cite{couetoux2011continuous}. 
    \item We show through experiments that the proposed technique achieves better or equal results by using significantly shallower quantum circuits, less classical and quantum computational resources compared to the best previous work based on MCTS \cite{wang}.
    \item We extend the study to an unstructured problem \cite{lu2023qas}. We generate random quantum circuits of various sizes and evaluate the performance of designing PQCs that approximate these target circuits. The distance between the quantum circuits is measured in terms of quantum fidelity \cite{nielsen2010quantum}. Furthermore, we analyze the ability of our MCTS approach to find classically hard-to-simulate quantum circuits, measured in terms of their stabilizer R\'enyi entropy $M_2$ \cite{leone2022stabilizer}.
\end{enumerate}
PWMCTS improves upon the previous MCTS work for QAS \cite{wang} on quantum and classical computational resources and automation, namely the number of quantum circuit evaluations, the number of quantum gates (especially CNOTs), and the number of hyperparameters to tune.

This article is organized as follows. First, Section \ref{related_works} provides a comprehensive overview of existing approaches and advancements in the field of QAS. Next, Section \ref{methods} provides a full description of PWMCTS. Section \ref{applications} introduces all the applications, including the experimental setup and results. Section \ref{discussion} examines the results with a comparison to related works. Finally, we highlight the strengths and weaknesses of our technique with an outlook for future research directions in Section \ref{conclusions}.

\section{Related Work} \label{related_works}
Neural Architecture Search (NAS) is the precursor of Quantum Architecture Search (QAS) within the domain of classical machine learning, which aims to automate the process of finding the architecture of artificial neural networks. According to these similarities, Zhang et al. \cite{zhang2022differentiable} proposed a differentiable quantum architecture search inspired by the differentiable neural architecture search developed for NAS \cite{liu2018darts}. The same authors also developed a neural predictor-based method \cite{zhang2021neural}. In this scenario, Du et al. \cite{du2022quantum} proposed an algorithm composed of two main parts. First, it initializes a set of PQCs, called supernet, by uniformly sampling them within the constraint of a fixed number of qubits, types of gates, and maximum circuit depth. The supernet is equipped with a weight-sharing strategy and then trained. Second, all the circuits of the supernet are ranked in terms of performance and the best are fine-tuned. Notably, this algorithm has been tested through both simulations and real experiments. 

\noindent Although QAS and NAS are closely related, the quantum and classical models can substantially differ in generalization and fundamental structure. A meta-learning approach has also been proposed, involving the discovery of a meta-heuristic that combines topology and parameters, followed by an adaptation phase for new tasks. \cite{he2022quantum}.

Evolutionary algorithms are also a promising approach to address QAS problems. The first attempt used genetic programming to find the quantum circuit implementing the quantum teleportation protocol \cite{williams1998automated}.  Li et al. \cite{li2017approximate} implemented a genetic algorithm to find an approximate quantum circuit for quantum adders. Zhang et al. developed \cite{zhang2023evolutionary} an evolutionary-based quantum architecture search assisted by the quantum Fisher information matrix to remove redundant parameters in PQCs. It has been tested on classification tasks of the MNIST dataset, with experiments with three and four qubits. Chivilikhin et al. \cite{chivilikhin2020mog} proposed a multi-objective approach, composed of a genetic algorithm and an evolutionary strategy working on the topology and parameters of the PQC respectively. It has been tested on variational quantum eigensolvers for several molecules. Evolutionary algorithms present typical advantages of population-based and derivative-free approaches, which help the search to avoid local minima. At the same time, they can be expensive in terms of computational resources on large problems. 
The topology of the searched PQC may have significant differences by varying the problem domain or even only changing instances in the same domain. However, in some applications, PQCs may present repeating patterns. It has been shown that a hierarchical representation of quantum circuits can benefit the search for genetic algorithms on quantum machine-learning problems \cite{lourens2023hierarchical}.

Reinforcement learning is also a promising approach for QAS, as shown by different frameworks developed for deep reinforcement learning agents \cite{niu2019universal, fosel2018reinforcement,  ostaszewski2021reinforcement, kuo2021quantum}. Those methods require a considerable amount of training data and computational resources. Moreover, deep reinforcement learning agents can easily get stuck in local minima as shown on the classical analogue NAS problems \cite{pham2018efficient}. Recent advances in reinforcement learning also considered the effect of quantum noise in the QAS for the ground-state energy problem on three benchmark molecules \cite{patel2024curriculum}.

Training-free approaches to QAS are also worth mentioning for the reduced computational cost required to evaluate quantum circuits. He et al. \cite{he2024training} proposed a method based on the directed acyclic graph representation of quantum circuits, where circuits are evaluated based on the number of paths in their graph representation and their expressibility. On one hand, this problem-agnostic method is computationally efficient. On the other hand, additional problem-dependent metrics are required for effective applications to VQAs. Expressibility of quantum circuits has also been recently used to pretrain a self-supervised learning model, which, when combined with Bayesian optimization, provides a robust solution for QAS \cite{he2025self}.
Motivated by the limitations of the NISQ devices, Situ et al. \cite{situ2024distributed} managed to develop a distributed approach to QAS, which also employs a training-free evaluation methodology. 

Inspired by \cite{alphazero}, NAS problems have also been formulated in a tree representation to exploit Monte Carlo Tree Search (MCTS) techniques to design deep neural networks \cite{wang2020neural, huang2021neural}. Analogously, Meng et al. \cite{meng2021quantum} proposed a tree representation of QAS and employed MCTS to design PQCs for quantum chemistry and condensed matter physics. Subsequently, Wang et. al. \cite{wang} have generalized the method, which achieves remarkable results on a wide range of applications, including the ground state energy problem for three different molecules, a system of linear equations, and more.  The problem is formulated in a tree structure and a Nested MCTS (NMCTS) is employed for the search. Quantum circuits are sliced into a fixed number of empty layers and NMCTS fills them by choosing between a pre-defined pool of operations. Each layer of the quantum circuit corresponds to a tree level. The branching factor is given by the number of combinations that can be realized with the pool of operations. In this framework, an epoch is composed of the search technique acting on the topology of PQCs and a classical gradient-based technique acting on the angle parameters. This technique requires an expensive parameter optimization step at each epoch. Additionally, it requires fixing several hyperparameters that strongly affect the solution, such as the number of layers, which implicitly determines the number of gates in the PQC. However, these quantities are generally unknown and problem dependent. Hence, an expensive hyperparameter tuning has to be performed on several quantities for any problem to find a suitable ansatz. 

This article addresses the key limitations encountered by MCTS approaches introducing a different tree-based action space formulation and some enhancements to the vanilla MCTS. We refer the reader to Section \ref{mcts_qc} for the details. Our proposed PWMCTS approach provides not only a higher level of automation but also significant improvements in the number of circuit evaluations required to design the PQC and its gate count (including CNOTs).

\section{Progressive Widening enhanced Monte Carlo Tree Search} \label{methods}
This section presents the Progressive Widening enhanced Monte Carlo Tree Search technique (PWMCTS) for the Quantum Architecture Search problem. In contrast to MCTS-based prior research \cite{meng2021quantum,wang}, we do not use gradient-based techniques during the search. This results in a gradient-free MCTS technique that designs both the topology and the parameters of the PQCs. Note that a gradient-based optimizer is employed for standard fine-tuning of the parameters of the final circuit designed by MCTS. 

The section is organized as follows. First, we formally define the QAS problem in Subsection \ref{qas}. Then, we provide a general introduction to MCTS in Subsection \ref{mcts}. Finally, we present a detailed description of PWMCTS in Subsection \ref{mcts_qc}.

\subsection{Quantum Architecture Search} \label{qas}
Quantum Architecture Search (QAS) aims to find a PQC that optimizes the given objective function. The problem encoded into the VQA framework specifies the objective function and consequently the number of qubits $n$ of the PQC. QAS involves designing the PQC topology, which consists of an ordered sequence of quantum gates $V$, their respective positions, and the corresponding angle parameters $\theta$. 

\begin{figure}[!ht]
    \centering
    \includegraphics[width=0.8\textwidth]{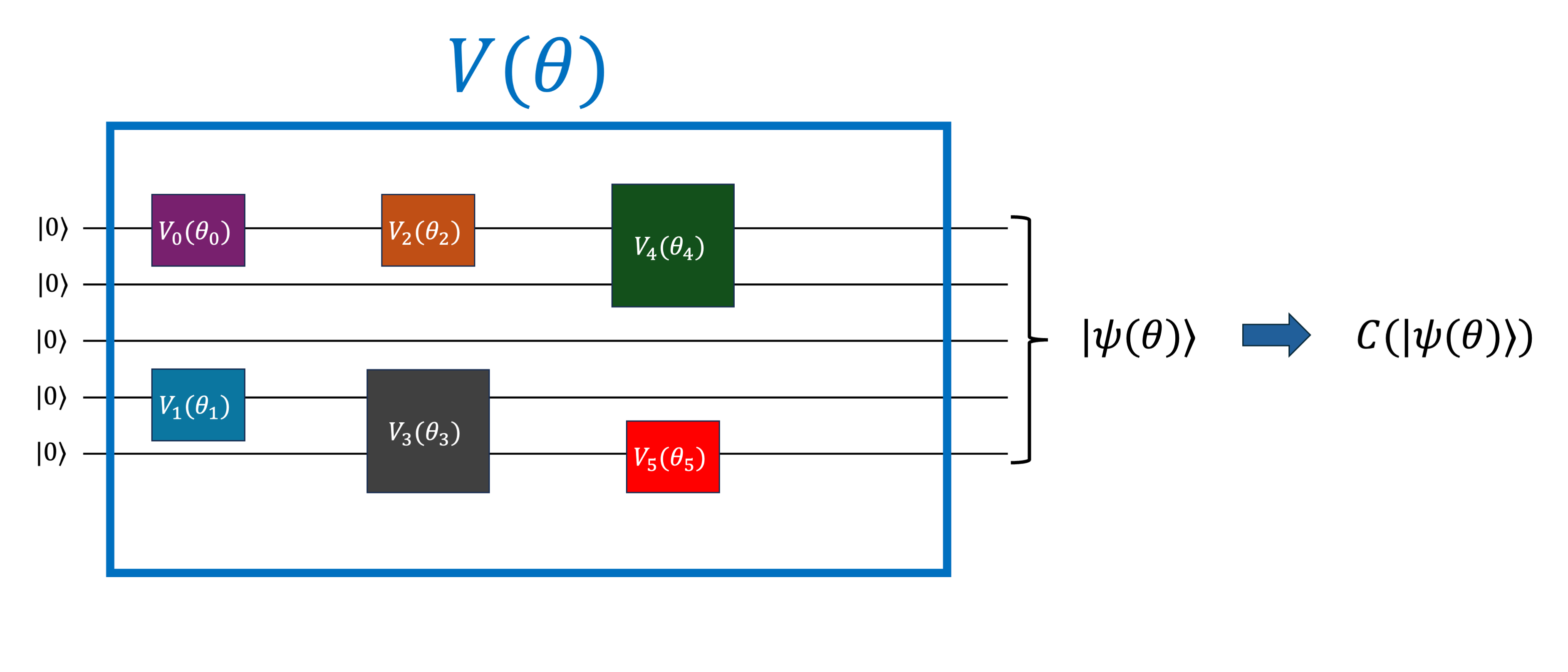}
    \caption{The cost function \( C \) is defined on the domain of variational quantum states \( \ket{\psi(\theta)} \). The goal in QAS is to design the unitary \( V(\theta) \) to prepare the state \( \ket{\psi(\theta)} = V(\theta)\ket{0} \), which minimizes the cost \( C(\ket{\psi(\theta)}) \).}
    \label{qas_scheme}
\end{figure}

The variational quantum state, $\ket{\psi(\theta)}= V(\theta)\ket{0}$, can be generally prepared by initializing all qubits to $\ket{0}$ and subsequently applying an ordered sequence of parameterized and non-parameterized gates (unitaries):
\begin{equation}
    V (\theta)= \prod_{i=1}^L V_i (\theta_i)  \, .\label{variational_circuit}
\end{equation}
The $n$-qubit unitary operator $V$ consists of an ordered sequence of $L$ quantum gates $V_i$, picked from a certain universal gate set. In this notation, the initial state is $\ket{0}=\ket{0}^{\otimes n }$ and $ V_i$ is either a single-qubit or $2$-qubit operator tensored with the identity operator on the subspace of the unaffected qubit(s). If $V_i$ is a non-parameterized gate, then $\theta_i$ is zero-dimensional. As a result, the vector of angle parameters is $\theta =(\theta_0,\theta_1, ..., \theta_l) \in [0,2\pi]^{l}$, with $l\leq L$. In the following, we will use the term \textit{variational quantum state} $\ket{\psi(\theta)}$ and \textit{quantum circuit} interchangeably. Figure \ref{qas_scheme} represents the QAS problem on a generic PQC.

The quality of a quantum circuit is measured by a real-valued objective function $C( \ket{\psi(\theta)})$, also referred to as \textit{cost} or \textit{loss} function. The original problem is encoded as a minimization task over the cost function $C$ defined on the domain of variational quantum states $\ket{\psi(\theta)}$. The best $\ket{\psi_0}$ corresponds to the minimum $C_0= C(\ket{\psi_0})$, 
\begin{equation}
    \ket{\psi_0} = \arg \min_{\ket{\psi(\theta)}} C(\ket{\psi(\theta) }\label{objective_function}.
\end{equation}

\subsection{Monte Carlo Tree Search}\label{mcts}
Monte Carlo Tree Search (MCTS) is a best-first search method that has achieved notable results on several search-based problems, especially in games \cite{browne2012survey, swiechowski,alphazero}. Its basic implementation does not require any domain-specific knowledge. This trait is beneficial for the design of algorithms for artificial general intelligence \cite{finnsson2012generalized,sironi2018self}. 

\noindent MCTS is based on a randomized exploration of the search space. Using the results of previous explorations, the algorithm gradually builds up a search tree in memory, and successively becomes better at accurately estimating the values of the most promising actions. It consists of four strategic steps \cite{browne2012survey}, repeated as long as there is time left. (1) In the \textit{selection} step the tree is traversed from the root node downwards until a state is chosen, which has not been stored in the tree. (2) Next, in the \textit{roll-out} step, actions are chosen in self-play until a terminal state is reached. (3) Subsequently, in the \textit{expansion} step one or more states encountered along its roll-out are added to the tree. (4) Finally, in the \textit{backpropagation} step, the result (reward) is propagated back along the previously traversed path up to the root node, where node statistics are updated accordingly.

MCTS usually starts with a tree containing only the root node, representing the initial state. The tree is gradually grown by executing the four steps described above. Each complete cycle of these steps is referred to as a full iteration.

\noindent On the one hand, MCTS aims to visit nodes that have led to the highest average reward so far (exploitation). On the other hand, it is crucial to explore less-visited nodes, as the uncertainty of simulations might mean they are closer to the optimum (exploration). The selection step chooses the child node to investigate based on previous information, balancing exploitation and exploration. Several selection strategies have been suggested for MCTS \cite{browne2012survey}, but the most popular one is based on the UCB1 algorithm \cite{auer2002finite}, called UCT (Upper Confidence Bounds applied to Trees) \cite{uct}.  
Given a state $s$ and the set of all possible actions $\mathcal{A}$, the UCT agent takes the action $a^*$ with the highest $UCB$ value
\begin{equation}
a^* = \arg \max_{a \in \mathcal{A}} \, UCB(s, a)   
\end{equation}
\begin{equation}
UCB(s, a) = \frac{Q_{(s,a)}}{N_{(s,a)}}+ c\, \sqrt{\frac{\ln N_s}{N_{(s,a)}}} \label{uct}
\end{equation}
where $N_{(s,a)}$ is the number of times the agent took the action $a$ from the state $s$, $N_s = \sum_{a\in \mathcal{A}} N_{(s,a)}$ is the total number of times the agent visited the state $s$ and $Q_{(s,a)}$ is the cumulative reward the agent gained by taking the action $a$ from the state $s$. The constant $c$ is a parameter that controls the degree of exploration, captured by the second term of Formula \ref{uct} versus the exploitation captured by the first term. The value $c$ can be tuned experimentally. 

\noindent The roll-out starts when the search reaches a leaf node. Hence, actions are selected in self-play until a terminal state is reached. This task might consist of playing plain random actions or, better, semi-random actions chosen according to a simulation strategy. In our agnostic approach, we work with a random roll-out. Simulation strategies based on heuristics have the potential to improve the performance of MCTS significantly \cite{winands2019monte}. 
\begin{figure}[!hb]
    \centering
    \includegraphics[scale=0.7]{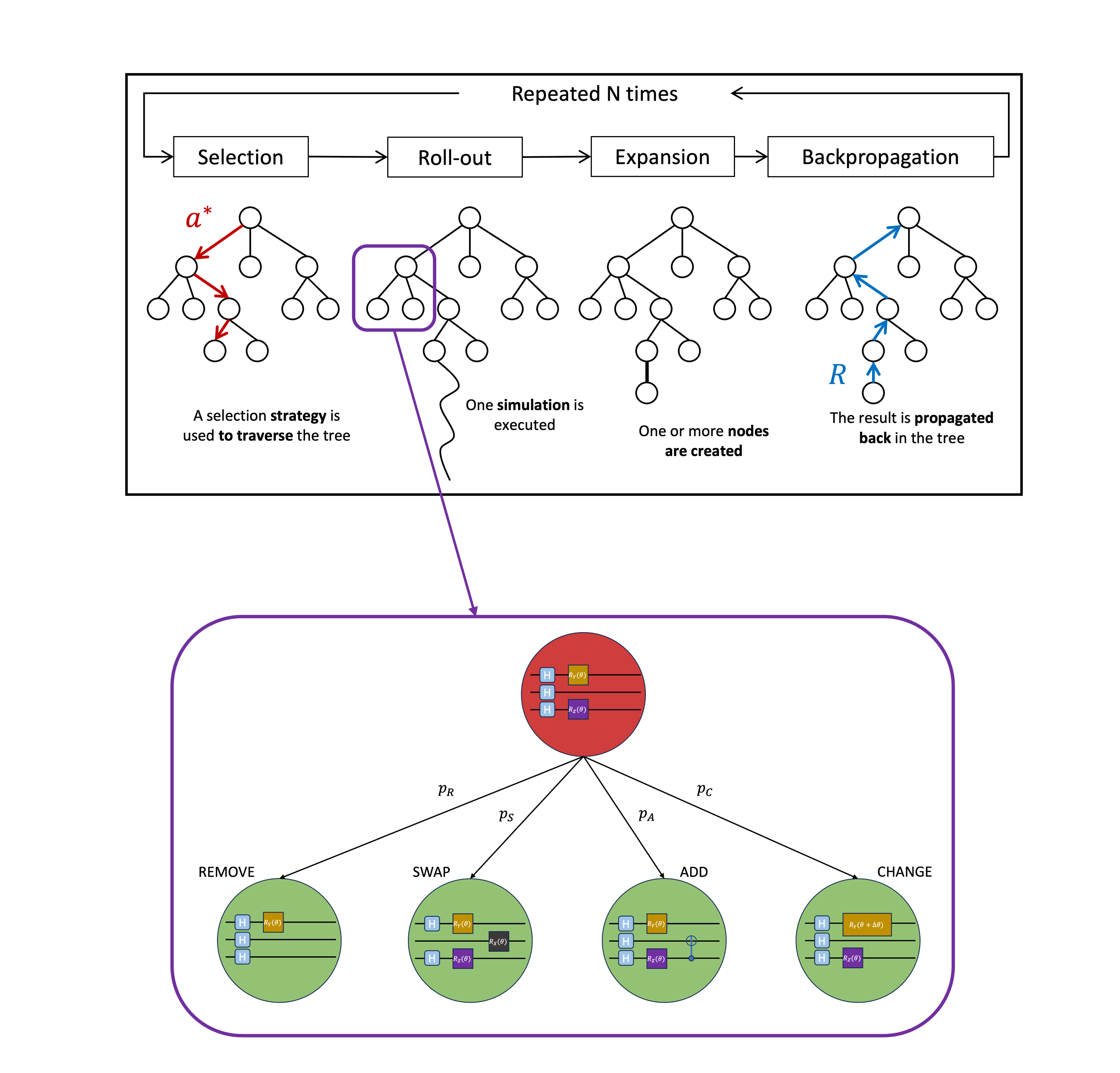}
    \caption{Monte Carlo Tree Search Scheme. Starting from the root node, the tree is gradually grown by executing the selection, roll-out, expansion, and backpropagation steps \cite{winands2019monte}. Our QAS framework defines the action space by sampling in an infinite set of possible quantum circuit modifications.}
    \label{mcts_scheme}
\end{figure}

\subsection{Monte Carlo Tree Search for Quantum Architecture Search}\label{mcts_qc} 

PQCs naturally define a continuous search space when considering real-valued angle parameters. Then, we first define the action space using a sampling technique for PQCs, borrowed from a quantum evolutionary strategy, which discretizes the space \cite{lukac2003evolutionary}. Second, we equip our MCTS with a progressive widening technique to search within the infinite number of actions \cite{couetoux2011continuous}. 

We propose a Progressive Widening enhanced MCTS (PWMCTS) placed in a framework where each node corresponds to an $n$-qubit quantum circuit and each action corresponds to a particular circuit modification. 
The reward for PWMCTS is a real-valued function $R$ in the domain of $n$-qubit quantum circuits reaching its maximum value for the circuit's optimal configuration. The objective function $C$ introduced in Section \ref{qas} is mapped onto the reward function through a problem-specific function $f$
\begin{equation}
    R (\ket{\psi (\theta)}) = f(C(\ket{\psi (\theta)})).
\end{equation}
In contrast to the domain of games, where the reward functions are typically defined for terminal states only, in our problem formulation the reward function is well-defined for all the tree nodes. Since each is potentially a terminal state, the roll-out step is not necessary to evaluate a newly visited node. Nevertheless, it may give insights into future actions during the search.

The initial circuit saved in the root node can be user-defined. By default, we start from the non-classical state generated by the quantum circuit composed by a Hadamard gate applied to each qubit, all initialized to $\ket{0}$.
PWMCTS explores the search space by sampling from four classes of allowed actions:

\begin{enumerate}
    \item \textit{adding} a random gate on a random qubit at the end of the circuit  (A);
    \item \textit{swapping} a random gate in the circuit with a new one (S);
    \item \textit{deleting} from the circuit a gate at a random position (D);
    \item \textit{changing} the value of the parameter $\theta_i$ of a randomly chosen parameterized gate (C). The new value is $\theta_i+\epsilon$, where $\epsilon \sim \mathcal{N}(0, \Delta \phi)$ is sampled from a normal distribution with mean zero and standard deviation $\Delta \phi$.
\end{enumerate}

PWMCTS chooses between the four classes of actions by sampling from a probability mass distribution 
\begin{equation}
    p=(p_A, p_S, p_D, p_C)
\end{equation}
where $p_A$, $p_S$, $p_D$, $p_C$ correspond to the probabilities of choosing the respective actions. This sampling technique is inspired by a framework proposed by Franken et al. \cite{franken2022quantum} designed for an evolutionary strategy on quantum circuits. Figure \ref{mcts_scheme} gives an overview of our approach. Note that $p$ is a hyperparameter of the model, which generally has to be tuned. However, at the beginning of the search, the circuit needs to be built up, then we set the initial sampling distribution to
\begin{equation}
    p_0 = (1, 0, 0, 0)
\end{equation}
which only allows the addition of new gates. The sampling probability is switched to $p$ as soon as the number of gates is double the number of qubits at least in one quantum circuit in the tree. Finally, there is a hardware-dependent control that modifies the probability $p$ when a condition is reached. In our implementation, the probability $p_A$ is fixed to zero when a certain circuit depth is reached. The user can set the above-mentioned condition to the number of CNOTs.

\noindent Quantum gates are sampled from the universal gate set $G$ consisting of the controlled-NOT gate and the parameterized single-qubit rotations \cite{nielsen2010quantum}. In general, $G$ can be set according to the specifics of the quantum hardware at disposal.
\begin{equation}
    G = ( \text{CNOT}, R_x, R_y, R_z) \label{gate_pool}
\end{equation}
Our formulation of the QAS problem works on a continuous action space as it deals with the position and type of gates as well as with their respective real-valued angle parameters. The sampling scheme on PQCs described above discretizes the action space, which is still infinite. The number of actions, the \textit{branching factor}, has to be finite and by fixing its value we get a hyperparameter of the model.

Preliminary experiments on the value of the branching factor of the tree show that it highly impacts the performance of MCTS in our QAS formulation, see Figure \ref{pw_fig}. For this purpose, we equipped MCTS with a progressive widening technique  \cite{chaslot, couetoux2011continuous}. It adaptively chooses the branching factor $k_s$ for each node, encoding the state $s$, in relation to the number of times the node gets visited $N_s$, according to the following formula
\begin{equation}
    k_s=\lceil \beta N_s^\alpha \rceil    \label{progressive_widening}
\end{equation} 
where $\beta >0$ and $\alpha \in ]0,1[$. We refer the reader to Figure \ref{pw_fig} for a better understanding of the impact of those hyperparameters on the branching factor and to Table \ref{table_hyperparameters} for the values used in our experiments.

\begin{table}[!b]
\caption{Hyperparameters with their respective symbols and values used throughout the article and the experiments. The value for $I$ is missing as it is explored in the experiments in Section \ref{applications}. }

\label{table_hyperparameters}
\centering
\begin{tabular}{@{}lll@{}}
\toprule
\textbf{Hyperparameter} & \textbf{Symbol} & \textbf{Value} \\
\midrule
Monte Carlo iterations               & $I$              & -   \\
Roll-out steps                        & $r$              & 0   \\
Action-by-action                      & $\rho$           & 5\% \\
UCB value                             & $c$              & 0.4   \\
Progressive widening coefficient      & $\alpha$         & 1 \\
Progressive widening exponent         & $\beta$          & 0.3   \\
Adding probability                    & $p_A$            & 0.5 \\
Swapping probability                  & $p_S$            & 0.2 \\
Changing probability                  & $p_C$            & 0.2 \\
Deleting probability                  & $p_D$            & 0.1 \\
Angle deviation in (C)              & $\Delta \phi$  & 0.2 \\
Maximum quantum circuit depth         & $d$              & 20  \\
Maximum Adam optimizer steps          & $T$              & 500 \\
\bottomrule
\end{tabular}
\end{table}

\noindent Because we are interested in the whole sequence of gates an action-by-action search is employed to distribute the search time along all levels of the tree \cite{baier2012nested, schadd2012single}.
Once a level of the tree is sufficiently explored, PWMCTS commits to the action that has the best score so far. Specifically, the PWMCTS agent commits to a node once its number of visits reaches a certain percentage, $\rho$, of the total number of Monte Carlo iterations, $I$. Then, the search continues starting from the node it committed to. Note that $\rho$ is a hyperparameter that affects the depth of the tree and consequently the number of gates in the quantum circuit.

\noindent At the end of the search, the \textit{best path} is retrieved as follows. The action $a$ with the highest cumulative reward $Q_{(s, a)}$ is taken in state $s$ starting from the root until a leaf node. The path retrieved represents the list of candidate quantum circuits as ansatz choice for the problem. Note that, even if the quality of quantum circuits is expected to improve along the paths from the root to the leaves, there is no guarantee that the cost will vary monotonously along these paths. For this reason, each quantum circuit belonging to the best path is evaluated one more time, and the best of them is chosen as ansatz.

Finally, the standard fine-tuning of the angle parameters of PQC found by PWMCTS is performed via the Adam optimizer \cite{kingma2014adam}. It is a gradient-based classical optimizer, chosen for its robustness and adaptability. It adjusts the learning rates adaptively for each parameter by computing the first and second moments of the gradients. We used the parameter-shift rule to apply a gradient descent technique in the circuit-based quantum computing setting \cite{wierichs2022general}. In the experiments, we use the Adam optimizer in its default PennyLane implementation \cite{bergholm2018pennylane}.

Table \ref{table_hyperparameters} gives an overview of the hyperparameters of PWMCTS and the values used in the experiments. The values of the single configuration of hyperparameters used in in this article have been determined using a grid search around the reference values found in the literature \cite{wang,franken2022quantum, couetoux2011continuous, sironi2019monte}. Note that, a further hyperparameter exploration may improve the results, especially on the progressive widening and action probability hyperparameters.

\section{Applications} \label{applications}
In this section, we discuss the application domains selected to test the PWMCTS technique. First, we present the structured problems, including the ground-state energy problem on three different molecules using VQE \cite{kandala2017hardware}, Section \ref{vqe}, and solving systems of linear equations via VQLS \cite{bravo2023variational}, Section \ref{vqls}. Second, we present the unstructured problem of quantum oracle approximation, which consists of approximating target quantum states defined in a dataset of random quantum circuits \cite{lu2023qas}, Section \ref{oracles}. In each section, we introduce the problem, the experimental setup, and the related numerical results.

\subsection{Variational Quantum Eigensolver} \label{vqe}
Ground state energy estimation is one of the most promising applications for quantum computing, with an expected exponential advantage over classical techniques \cite{dalzell2023quantum,lee2023evaluating}. This has a major impact on the field of quantum chemistry for the study of molecules. The problem is fully described by the Hamiltonian $H$ of the system. It takes into account the geometric structure of the molecule, given by the different atoms, their reciprocal positions, and the classical and quantum correlations existing between the elementary units composing that molecule.

The Variational Quantum Eigensolver (VQE) is a NISQ approach to the ground-state energy problem \cite{kandala2017hardware}. The objective is to minimize the cost function $C$ defined by the expectation value of the Hamiltonian $H$ of the system on the variational state $\ket{\psi} = \ket{\psi(\theta)}$:
\begin{equation}
    C(  \ket{\psi (\theta)}) = \bra{\psi} H \ket{\psi}
\end{equation}
it physically represents the energy of the system in the state $\ket{\psi (\theta)}$. The minimum value $C_0$ corresponds to the ground-state energy
\begin{equation}
    C_0 = \min_\psi \bra{\psi} H \ket{\psi}
\end{equation}
The reward function is defined as
\begin{equation}
    R(  \ket{\psi (\theta)}) = - C(  \ket{\psi (\theta)}).
\end{equation}
Note that, finding an ansatz with a 'good' overlap with the target quantum state is a fundamental requirement for many near-term quantum algorithms that aim to provide a greater than polynomial speedup \cite{dalzell2023quantum}.

\subsubsection{Experimental Results}\label{vqe_results}
The set of Hamiltonians used for all the experiments in this section have been generated from the quantum chemistry module in PennyLane \cite{bergholm2018pennylane} together with the electronic structure package PySCF. The Hamiltonian $H_\mu$ of a molecule $\mu$ is defined by fixing the atoms that compose it with their respective positions in the $3$-dimensional Cartesian coordinate system, the number of active electrons, and the number of active orbitals. 
All the numerical results are provided in atomic units (au) for the lengths and Hartree (Ha) for the energies.
We executed the algorithm $10$ times for each experiment to collect statistics on independent events. 

The results are compared with two classical benchmarks, the Self-Consistent Field (SCF) and the Full Configuration Interaction (FCI) methods. 
The SCF method is a mean-field approach that approximates the ground state using a single Slater determinant. It is a common, computationally efficient benchmark.
The FCI method provides an exact solution to the electronic problem within a given basis set. It captures all possible electronic correlations, serving as the standard for benchmarking quantum chemistry methods. However, it comes with a factorial-scaling computational cost. We refer the reader to Section \ref{comparison} for a comparison with related quantum approaches.

\begin{figure}[ht]
    \centering
    \begin{minipage}{0.6\textwidth} 
        \centering
        \begin{subfigure}{\textwidth} 
            \centering
            \includegraphics[width=0.6\textwidth]{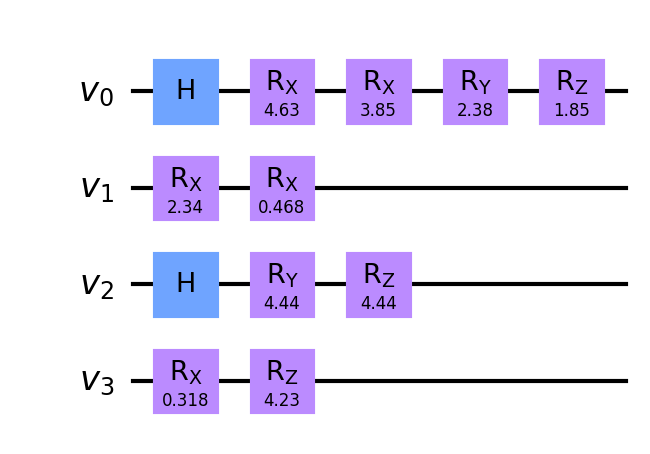}
            \caption{H$_2$}
            \label{h2_circuit}
        \end{subfigure}
        \vspace{0.5em} 
        \begin{subfigure}{1\textwidth} 
            \centering
            \includegraphics[width=0.75\textwidth]{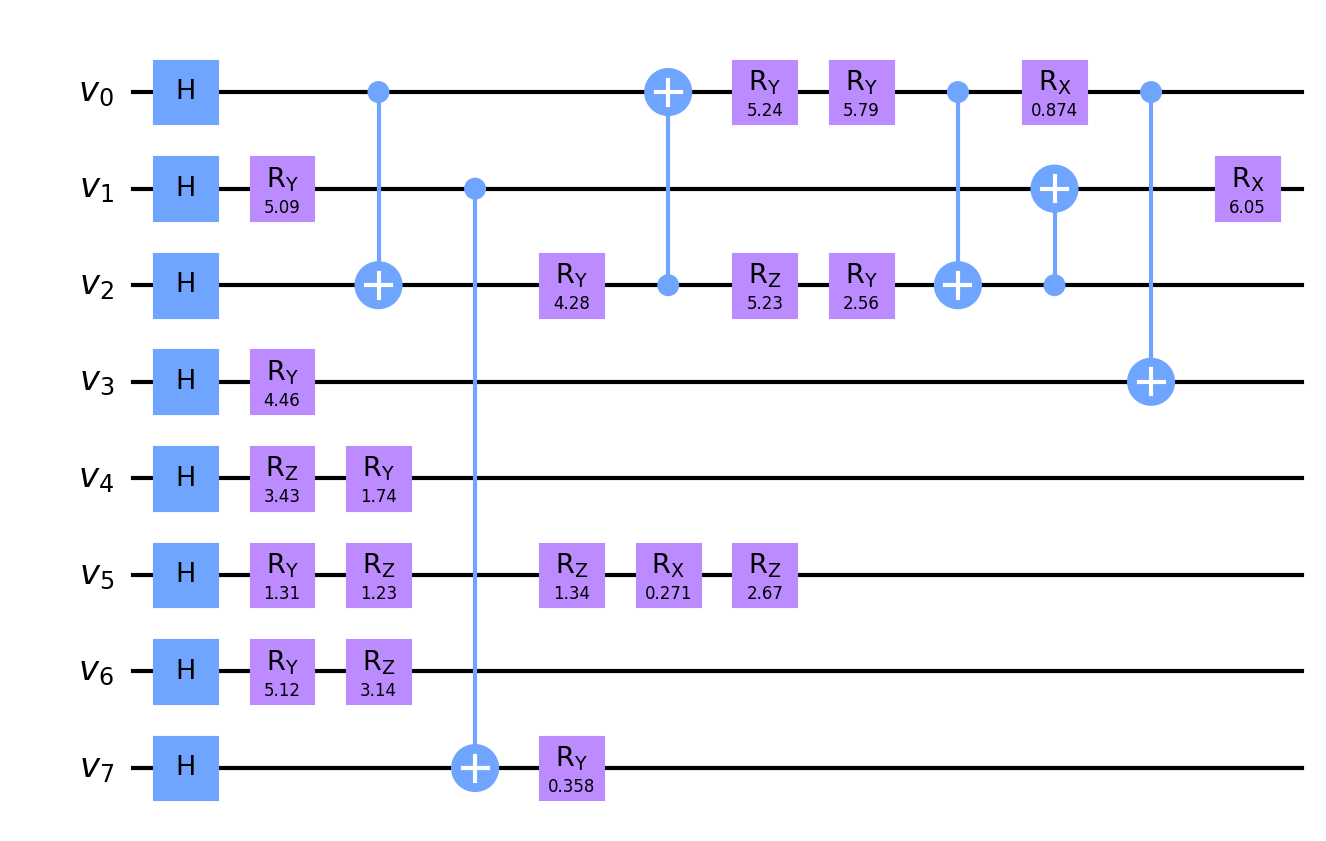}
            \caption{H$_2$O}
            \label{h2o_circuit}
        \end{subfigure}
    \end{minipage}%
    \hfill
    \begin{minipage}{0.39\textwidth} 
        \centering
        \begin{subfigure}{\textwidth}
            \centering
            \includegraphics[width=0.9\textwidth]{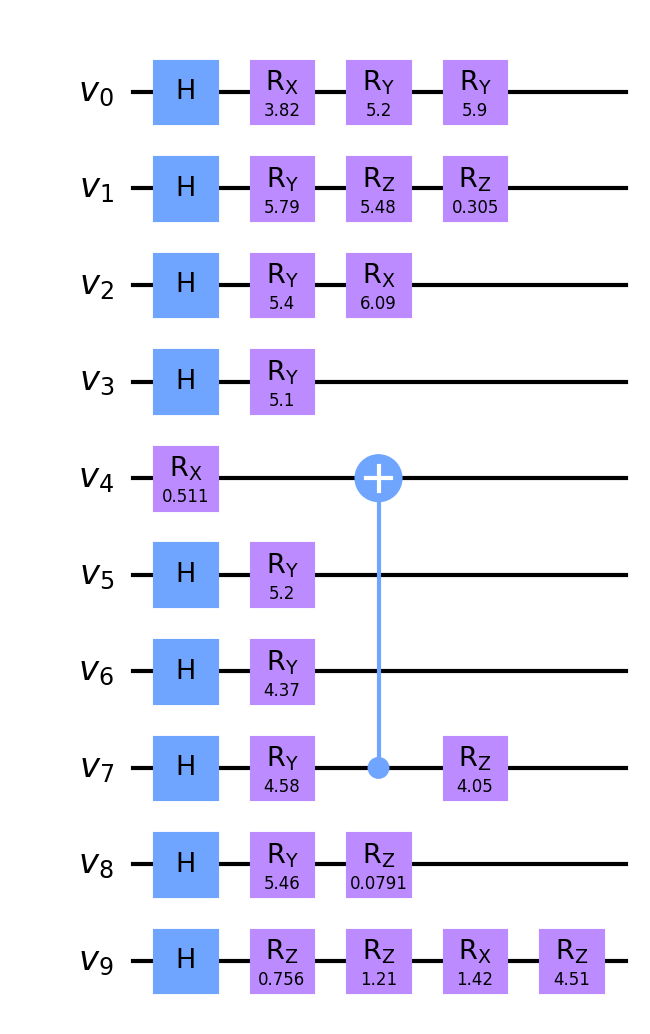}
            \caption{LiH}
            \label{lih_circuit}
        \end{subfigure}
    \end{minipage}
    \caption{PQCs designed by PWMCTS for the ground-state energy problem. The number of PWMCTS iterations $I$ is $1000$ for the molecule of Hydrogen (\ref{h2_circuit}), and $5000$ for the molecules of Water (\ref{h2o_circuit}) and Lithium Hydride (\ref{lih_circuit}).}
    \label{quantum_circuits}
\end{figure}

In the following, we present the experimental results obtained for the ground-state problem for the molecule of hydrogen H$_2$, lithium hydride LiH, and water H$_2$O. The PQCs provided by PWMCTS for the three ground-state problems are depicted in Figure \ref{quantum_circuits}. They all exhibit favorable characteristics for execution on NISQ devices, particularly a low CNOT count.
 
\begin{figure}[!ht]
\begin{subfigure}{.48\textwidth}
  \centering
  \includegraphics[width=1\textwidth]{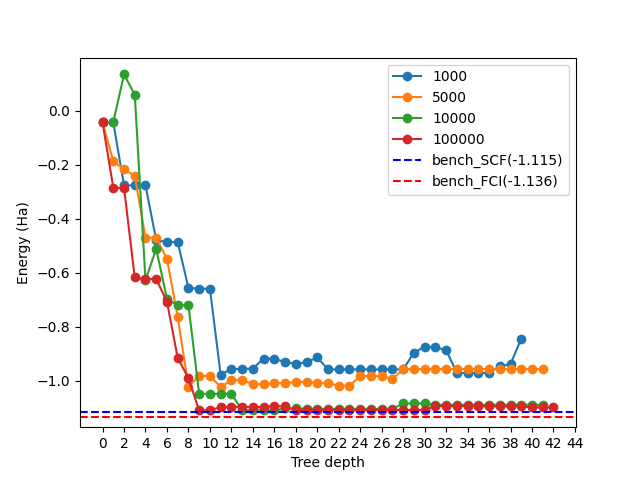}
  \caption{Cost along the best tree path}
  \label{h2_a}
\end{subfigure}
\begin{subfigure}{.48\textwidth}
  \centering
  \includegraphics[width=1\textwidth]{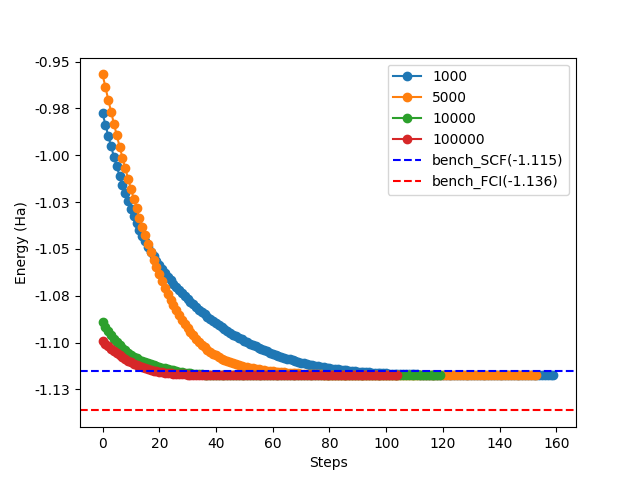}
  \caption{Fine-tuning of the best solution}
  \label{h2_b}
\end{subfigure}

\caption{H$_2$: Figure \ref{h2_a} plots the energy value along the best path retrieved by PWMCTS for different values of $I$.  Figure \ref{h2_b} shows the effect of the classical Adam optimizer on the angle parameters for different values of PWMCTS iterations. PWMCTS only requires $1000$ iterations to design an ansatz converging to the energy value $-1.117$ Ha after $160$ steps of the Adam optimizer on the angle parameters.}
\label{h2_fig}
\end{figure}

\subsubsection*{Ground State Energy of H$_2$}
In the three-dimensional space, the hydrogen molecule under study has two hydrogen atoms placed with coordinates $(0,0, -0.6614)$ and $(0,0, 0.6614)$. The number of active electrons, as well as the number of active orbitals, is $2$. The number of qubits required is $4$.

\noindent Figure \ref{h2_fig} shows the numerical results of PWMCTS on the ground-state problem for H$_2$. It converges to an approximate solution with only $1000$ iterations and less than $200$ iterations with the Adam Optimizer on the angle parameters. Increasing the value of $I$, PWMCTS designs PQCs that require fewer steps of the classical optimizer to fine-tune the parameters. Note that it does not obtain a result as good as the FCI method, but it improves upon the SCF method. Compared to related quantum approaches it reaches the same local minima using considerably fewer quantum circuit evaluations, see Table \ref{table_comparison}.

\subsubsection*{Ground State Energy of H$_2$O}
The water molecule studied has the two hydrogen atoms in $(0,0, 0)$ and $(3.3609,0, 0)$, while the oxygen atom in $(1.6323,0.8641, 0)$. The numbers of active electrons and active orbitals are both set to $4$. The number of qubits required is $8$.

\begin{figure}[!t]
\begin{subfigure}{.48\textwidth}
  \centering
  \includegraphics[width=1\textwidth]{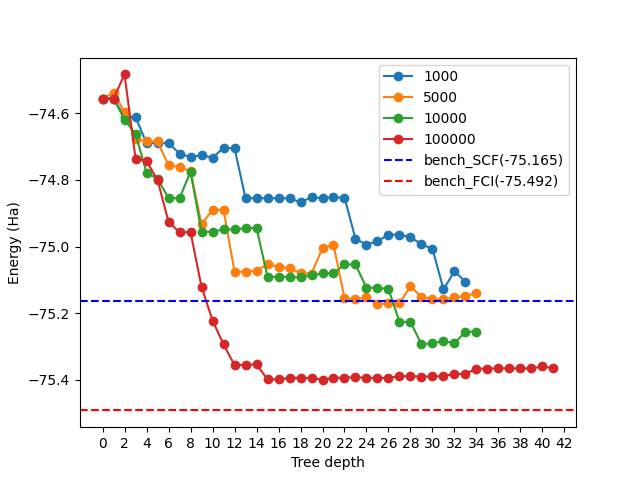}
  \caption{Cost along the best tree path  }
  \label{h2o_a}
\end{subfigure}
\begin{subfigure}{.48\textwidth}
  \centering
  \includegraphics[width=1\textwidth]{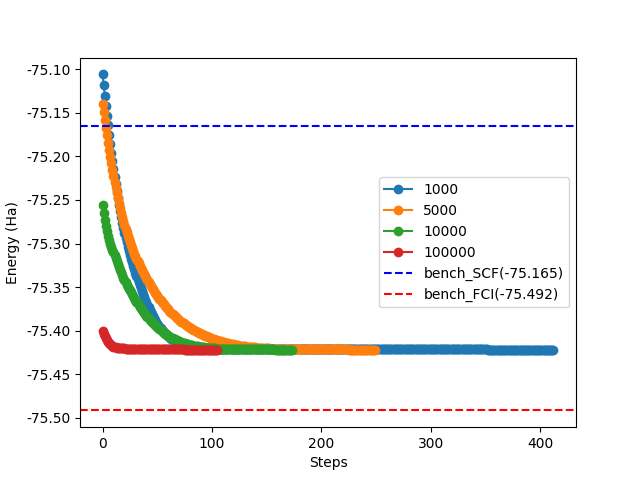}
  \caption{Fine-tuning of the best solution}
  \label{h2o_b}
\end{subfigure}
\caption{H$_2$O. Figure \ref{h2o_a} shows the energy value along the best path retrieved by PWMCTS and Figure \ref{h2o_b} the effect of the fine-tuning of the parameters. The results relate to the best result over $10$ independent runs and for different values of $I$. After fine-tuning, PWMCTS achieves the energy $-75.42$ Ha for all budget values.}
\label{h2o_fig}
\end{figure}

\noindent Figure \ref{h2o_fig} shows the numerical results of PWMCTS on the ground-state problem for H$_2$O. Fixing $I\geq 5000$, PWMCTS achieves lower energies than the SCF method even without fine-tuning the parameters. After fine-tuning, it converges consistently to the same value.

\begin{figure}[!b]
\begin{subfigure}{.48\textwidth}
  \centering
  \includegraphics[width=1\textwidth]{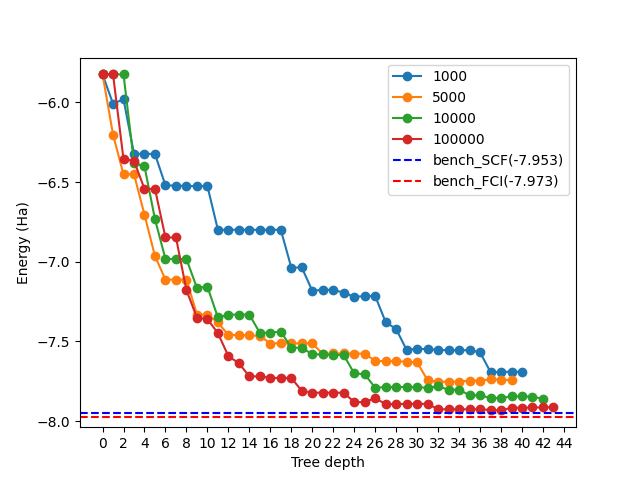}
  \caption{Cost along the best tree path  }
  \label{lih_a}
\end{subfigure}
\begin{subfigure}{.48\textwidth}
  \centering
  \includegraphics[width=1\textwidth]{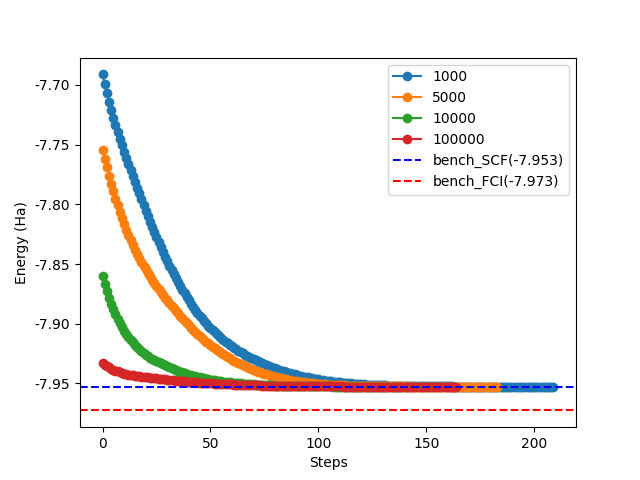}
  \caption{Fine-tuning of the best solution}
  \label{lih_b}
\end{subfigure}
\caption{LiH. Figure \ref{lih_a} shows the energy value along the best path retrieved by PWMCTS and Figure \ref{lih_b} the effect of the fine-tuning of the parameters. The results relate to the best result over $10$ independent runs and for different values of $I$. PWMCTS consistently estimates the energy of LiH to -7.9526 Ha.}
\label{lih_fig}
\end{figure}

\subsubsection*{Ground State Energy of LiH}
The Lithium hydride under study has a lithium atom in  $(0,0, 0)$  and a hydrogen atom in $(0, 0, 2.969280527)$. The number of active electrons and active orbitals is set to $2$ and $5$, respectively. The number of qubits required is $10$.

\noindent Figure \ref{lih_fig} shows the numerical results of PWMCTS on the ground-state problem for LiH. It converges to energies better than SCF with a $I\geq 5000$ before the fine-tuning. After fine-tuning the parameters of the PQC designed, PWMCTS consistently improves the SCF results on LiH, see Figure \ref{lih_b}.

\subsection{Variational Quantum Linear Solver}  \label{vqls}
Systems of linear equations promise to be the breeding ground for quantum computing as shown by the Harrow–Hassidim–Lloyd algorithm \cite{harrow2009quantum}. It provides an exponential speedup compared to classical algorithms under certain assumptions. However, as it makes use of quantum phase estimation it cannot be largely implemented in NISQ devices.

Bravo-Prieto et al. \cite{bravo2023variational} proposed the variational quantum linear solver (VQLS) to solve systems of linear equations $A x=b$ with a variational approach. They proved its efficient scaling in the linear system size, in the condition number of the matrix $A$, and in the error tolerance.

\noindent The problem is defined by a matrix $A\in \mathbb{C}^{2^n\times 2^n}$ and an $n$-qubit quantum state $\ket{b}$ that can be efficiently prepared through quantum circuits. The quantum state is $\ket{b}=U\ket{0}$, with $U=H^{\otimes n}$, while the matrix $A$ is a complex linear combination of unitaries $A_m$ that are easy to prepare:
\begin{equation}
    A=\sum_{m=0}^{M-1} c_mA_m\label{vqls_matrix}
\end{equation}
where $M$ has to be a polynomial function of the number of qubits $n$. The goal is to find the solution $x$ encoded in the amplitudes of the quantum state $\ket{x}$ such that $A\ket{x}\propto \ket{b}$.
The quality of the solution provided by the variational quantum state $V\ket{0}$ is measured by the local cost function $C$ for VQLS defined below.
\begin{equation}
    C = 1-\frac{\sum_{m,m'}c_mc_{m'} \bra{0} V^{\dag} A_{m'}^{\dag} U P U^{\dag} A_m V\ket{0}}{\sum_{m,m'}c_mc_{m'} \bra{0} V^{\dag} A_{m'}^{\dag}  A_m V\ket{0}}
\end{equation}
where $P$ is the projector
$P=\frac{\mathbb{I}}{2} +\frac{1}{2n} \sum_{j=0}^{n-1}Z_j  $, and $Z_j$ is the Pauli operator applied to the $j$-th qubit \cite{bravo2023variational}. 
Intuitively, $C$ quantifies the orthogonal component of $A \ket{x}$ with respect to $\ket{b}=U\ket{0}$, where the solution $\ket{x}$ is replaced by the variational state $V\ket{0}$. The variational quantum state affects the cost function $C = C(\ket{\psi (\theta)})$ through $V$.  
The reward function for PWMCTS in this domain is defined as
\begin{equation}
    R(  \ket{\psi (\theta)}) =  e^{{-10 \,C(  \ket{\psi (\theta)})}}.
\end{equation}

\subsubsection{Experimental Results} \label{vqls_results}
The experiments have been carried out on the $4$-qubit problem defined by the following linear system. 
\begin{equation}
    A = \alpha_0 \mathbb{I}+ \alpha_1 X_1+ \alpha_2X_2+\alpha_3Z_3Z_4 \label{vqls_A}
\end{equation}
\begin{equation}
    \ket{b}= H^{\otimes 4} \ket{0}^{\otimes 4} \label{vqls_b_vector}
\end{equation}
with $\alpha_0=0.1$, $\alpha_1=\alpha_2=1$ and $\alpha_3=0.2$. 
\begin{figure}[!ht]
\begin{subfigure}{.48\textwidth}
  \centering
  \includegraphics[width=1\textwidth]{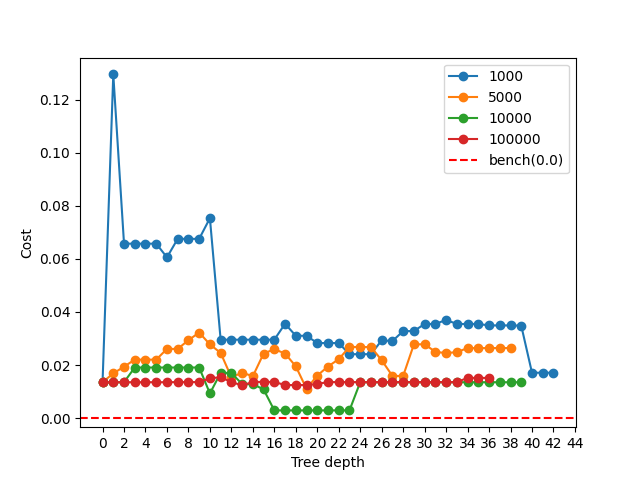}
  \caption{Cost along the best tree path  }
  \label{vqls_a}
\end{subfigure}
\begin{subfigure}{.48\textwidth}
  \centering
  \includegraphics[width=1\textwidth]{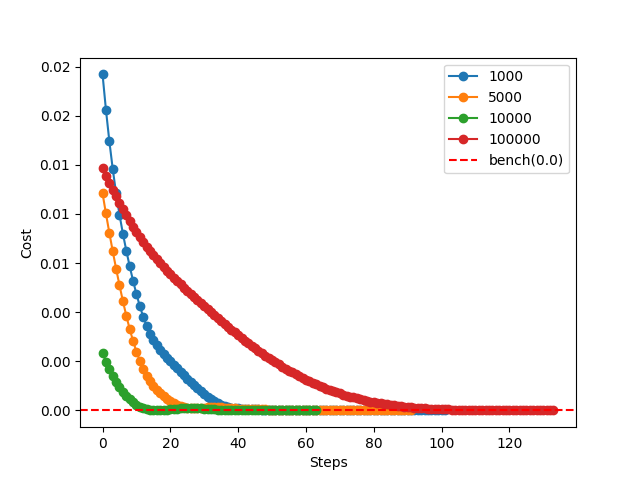}
  \caption{Fine-tuning of the best solution}
  \label{vqls_b}
\end{subfigure}
\caption{VQLS. Figure \ref{vqls_a} shows the cost value along the best path retrieved by PWMCTS and Figure \ref{vqls_b} the effect of the fine-tuning of the parameters. The results relate to the best cost value achieved over $10$ independent runs and for different values of $I$. PWMCTS achieves the notable cost value $3.98\times 10^{-4}$ using $I=10000$ iterations and $T=65$ steps of the Adam optimizer. }
\label{vqls_figure}
\end{figure}

The quantum circuits simulated in this application are $5$-qubit circuits, because of an ancilla qubit required by the VQLS algorithm.

\noindent In this application, the optimal cost value is zero and PWMCTS designs PQCs whose cost converges to very small values for different fixed iteration values, see Figure \ref{vqls}. PWMCTS achieves a cost value of $4.63 \cdot 10^{-7}$, $9.31 \cdot 10^{-8}$, $3.98 \cdot 10^{-8}$, and $1.83 \cdot 10^{-10}$ using $1000$, $5000$, $10000$, and $100000$ iterations $I$, respectively. By increasing the iterations $I$, that is the computational budget, the technique achieves better cost values. Figure \ref{vqls_a} shows the cost value over the best path retrieved by PWMCTS. In contrast to previous applications, PWMCTS struggles to find improvements in this domain, as most of the time it does not find an improved modification of the initial quantum circuit along the tree. However, Figure \ref{vqls_b} shows that the circuits designed are expressive enough to converge to the solution. 

\begin{figure}[t]
    \centering
    \includegraphics[width=0.6\linewidth]{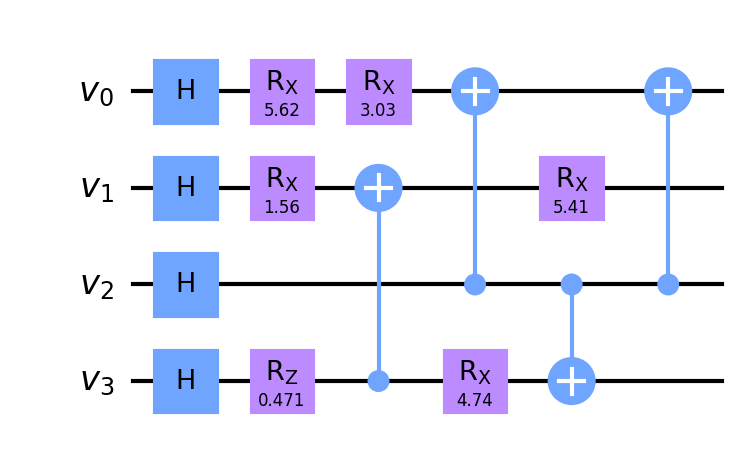}
    \caption{The PQC designed by PWMCTS with $I=10000$ to solve the system of linear equations.}
    \label{vqls_circuit}
\end{figure}

Figure \ref{vqls_circuit} shows the quantum ansatz provided by PWMCTS for the system of linear equations defined in Equations \ref{vqls_A} and \ref{vqls_b_vector}.

\subsection{Quantum Oracle Approximation}\label{oracles}
In this application, we test PWMCTS in an unstructured problem setting. A dataset of quantum circuits randomly generated defines the target solutions. The motivation for this application is twofold. On the one hand, we test our approach on a problem that is free from the underlying symmetries of particular domains. On the other hand, we analyze PWMCTS as a technique to design quantum circuits that are difficult to simulate on a classical computer. 

\noindent The classical simulability of quantum states is fundamental for assessing the achievable quantum advantage. 
In this context, nonstabilizerness, commonly known as quantum \textit{magic}, is a measure for quantum states that describes their difficulty to be characterized and manipulated on a classical computer \cite{nielsen2010quantum, bravyi2005universal,bartlett2014powered, howard2017appl}. 
For this reason, the target random quantum circuits are ranked according to their level of relevance based on their value of stabilizer R\'enyi entropy \cite{leone2022stabilizer}, chosen to measure the magic of quantum states due to its computational properties.

\noindent Given a pure quantum state $\ket{\psi}$, its $\alpha$-R\'enyi entropy is defined as
\begin{equation}
    M_\alpha (\ket{\psi} = \alpha(1-\alpha)^{-1} \log \sum_{P \in \mathcal{P}_n} \Xi^\alpha_P(\ket{\psi})  -\log 2^n
\end{equation}
where $\Xi_P= \frac{1}{2^n}\bra{\psi}P\ket{\psi}$ and $\mathcal{P}_n$ is the group of all the $n$-qubit Pauli strings with phases one \cite{leone2022stabilizer}. Specifically, we calculated $M_2$ for all the generated quantum states (circuits).

In the following, we describe the dataset generation. All quantum circuits are initialized to $\ket{0}^{\otimes n}$. Subsequently, we apply quantum gates sampled from the universal gate set composed of the Clifford generators (the Hadamard gate $H$, the phase gate $I$, the controlled-NOT) plus the non-Clifford $T$ gate.
The problem instances in this domain are defined by target quantum circuits that differ in the number of qubits, \( n \), number of gates, \( g \), and magic. We generate $10$ quantum circuits for each combination ($n$, $g$) and select the two of them characterized by the lowest and highest values of stabilizer R\'enyi entropy \( M_2 \). The number of qubits varies in the set \{4, 6, 8\}, the number of gates in \{5, 10, 15, 20, 30\} and the magic \{\textit{easy}, \textit{hard}\}. As a result, we have $30$ different target circuits labeled by $(n, g, M_2)$, each defining a different problem for PWMCTS.
Note that, when we refer to the hardest (easiest) quantum circuit generated, it is relative to the quantum circuits in our dataset, hence, it does not correspond to the overall hardest (easiest) quantum circuit to simulate classically given the qubits and gates constraints. 
The number of $T$ gates involved in creating a quantum state is directly related to its stabilizer R\'enyi entropy \cite{leone2022stabilizer}. However, high total quantum gates $g$ in the target quantum circuits do not guarantee high magic values. This is a consequence of the random process employed in the dataset generation. 
\noindent Note that, the universal gate set used for the data generation is different from that used by PWMCTS. This choice was made to avoid the undesired structural connection between the problem and the model, which could affect the fairness of the experiments.

In this application, the reward is defined by the quantum fidelity in its form for pure states \cite{nielsen2010quantum} 
\begin{equation}
    R (\ket{\psi (\theta)}) = |\braket{\phi|\psi (\theta)}|^2
\end{equation}
where $\ket{\phi}$ is the target state defined in the quantum circuit dataset and $R \in[0,1]$.
The cost function instead is defined as
\begin{equation}
    C (\ket{\psi (\theta)}) = 1 - R (\ket{\psi (\theta)})
\end{equation}
such that the minimum of the cost function is $C_0=0$.

\subsubsection{Experimental Results}\label{oracle_experiments}
 In this application, we have $30$ different target circuits labeled by the number of qubits, \( n \), number of gates, \( g \), and stabilizer R\'enyi entropy \( M_2 \). Each combination $(n, g, M_2)$ defines a different problem for PWMCTS. For each experiment $(n, g, M_2)$ we execute $10$ independent runs. 

\begin{figure}[!hb]
\begin{subfigure}{.48\textwidth}
  \centering
  \includegraphics[width=1\textwidth]{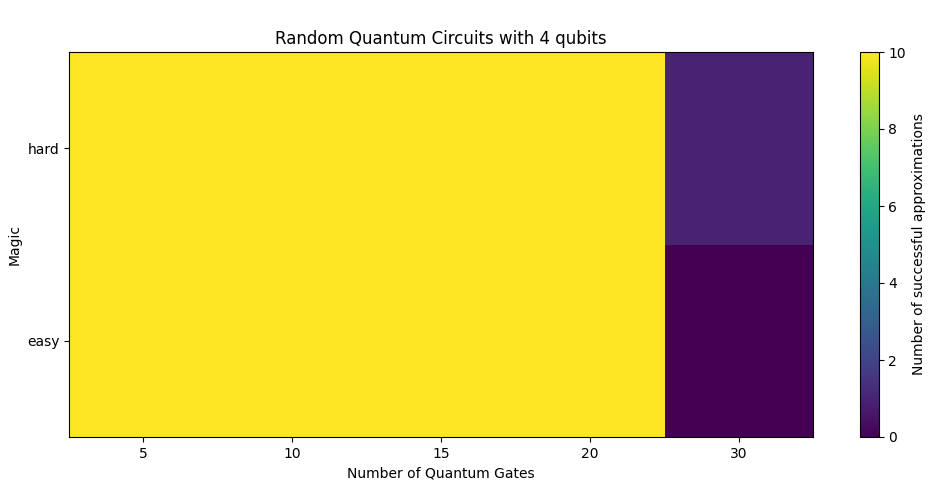}
  \caption{$4$ qubits, $I=100000$, $\epsilon = 0.05$}
  \label{oracle_4a}
\end{subfigure}
\begin{subfigure}{.48\textwidth}
  \centering
  \includegraphics[width=1\textwidth]{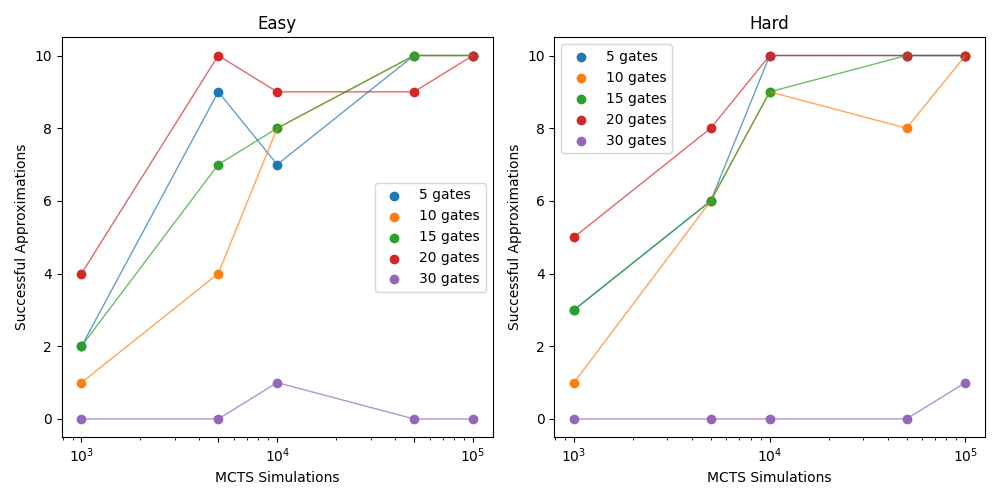}
  \caption{$4$ qubits, $\epsilon = 0.05$}
  \label{oracle_4b}
\end{subfigure}
\begin{subfigure}{.48\textwidth}
  \centering
  \includegraphics[width=1\textwidth]{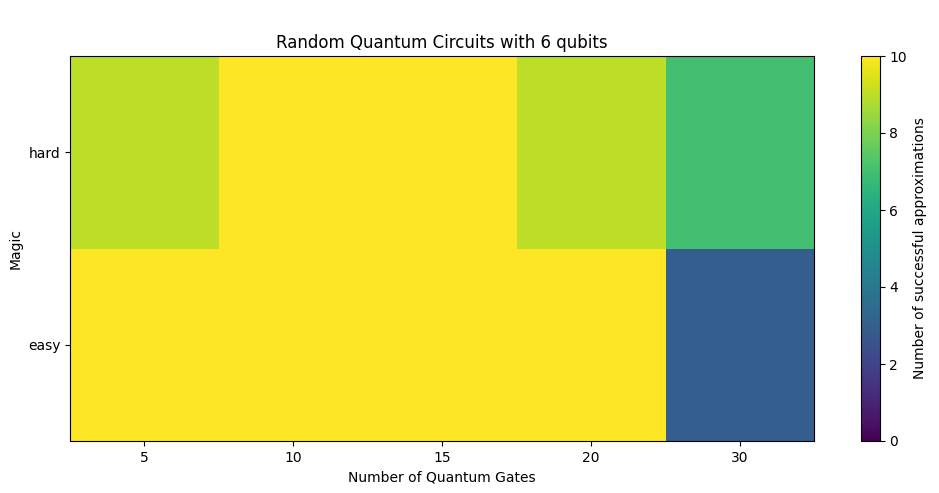}
  \caption{$6$ qubits, $I=100000$, $\epsilon = 0.1$}
  \label{oracle_6a}
\end{subfigure}
\begin{subfigure}{.48\textwidth}
  \centering
  \includegraphics[width=1\textwidth]{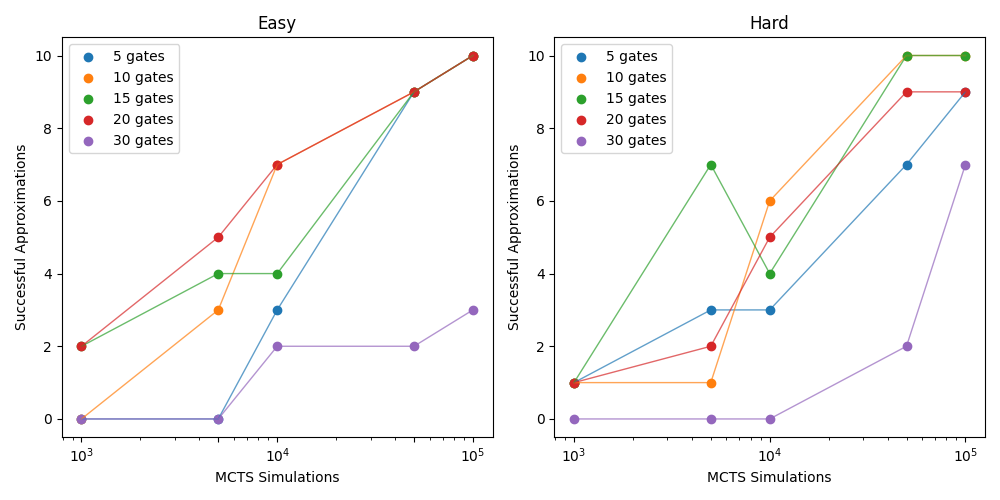}
  \caption{$6$ qubits, $\epsilon = 0.1$}
  \label{oracle_6b}
\end{subfigure}
\begin{subfigure}{.48\textwidth}
  \centering
  \includegraphics[width=1\textwidth]{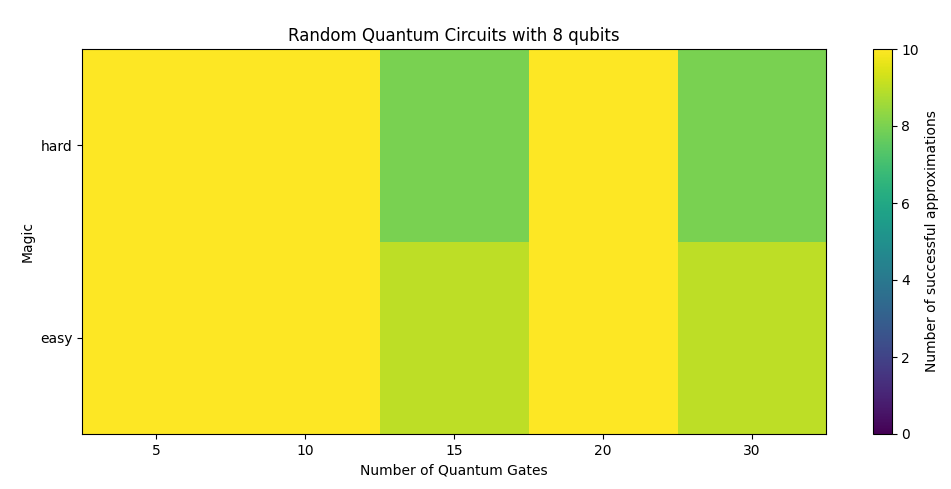}
  \caption{$8$ qubits, $I=100000$, $\epsilon = 0.2$}
  \label{oracle_8a}
\end{subfigure}
\begin{subfigure}{.48\textwidth}
  \centering
  \includegraphics[width=1\textwidth]{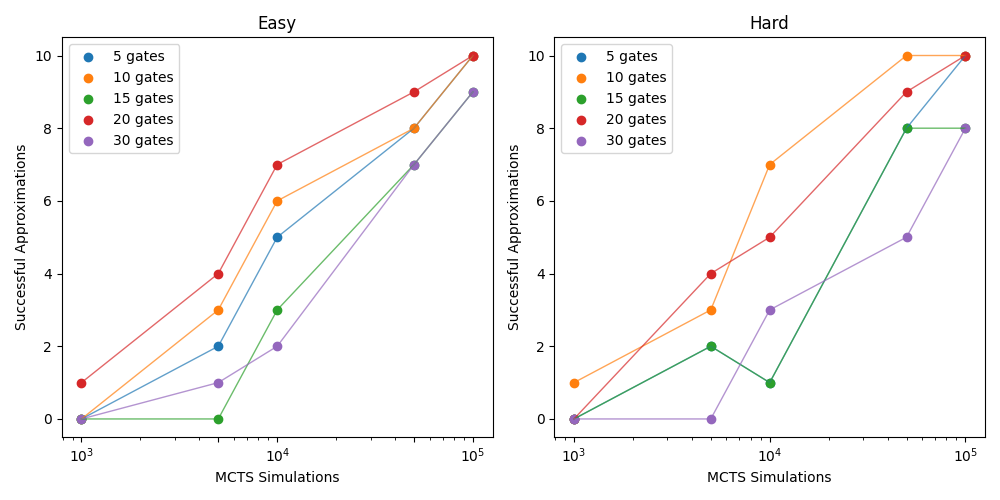}
  \caption{$8$ qubits, $\epsilon = 0.2$}
  \label{oracle_8b}
\end{subfigure}
\caption{Quantum oracle approximation. On the left, the color maps show the performance in terms of the number of successful approximations over independent runs with fixed $I$ and $\epsilon$, \ref{oracle_4a}, \ref{oracle_6a} and \ref{oracle_8a}. On the right, the plots show the impact of the number of PWMCTS iterations $I$ ($x$-axis in logarithmic scale) on the number of successful approximations ($y$-axis), \ref{oracle_4b}, \ref{oracle_6b} and \ref{oracle_8b}. }
\label{oracle_figure}
\end{figure}

\noindent We say that a variational quantum state $\ket{\psi}$ is an $\epsilon$-approximation for the quantum state $\ket{\phi}$ if 
\begin{equation}
    |\braket{\phi|\psi}|^2 \geq 1-\epsilon. \label{eps_approx}
\end{equation}
In our experiments, we consider the algorithm successful when it manages to approximate with $\epsilon=0.05$ the $4$-qubits circuits, $\epsilon=0.1$ the $6$-qubits circuits, $\epsilon=0.2$ the $8$-qubits circuits. The difficulty of the problem increases with the number of qubits. The choice of the values of $\epsilon$ and the size of quantum circuits allows to keep the number of PWMCTS iterations $I$ low and consequently the computation time. 

The performance of our technique improves by increasing the number of PWMCTS iterations in all experiments $(n, g, M_2)$, demonstrating the capabilities of our approach in this domain.
However, there are some cases where the algorithm cannot find good approximations. This is the case of the target $4$-qubits quantum circuits with $30$ gates.

\noindent On the one hand, the problem's difficulty is related to the number of qubits and the value of $\epsilon$. On the other hand, the number of gates used to generate the target circuit does not directly relate to the difficulty of the problem, which rather depends on the number of quantum gates needed to approximate the solution using the universal gate set $G$ used by PWMCTS, defined in Equation \ref{gate_pool}.
Furthermore, we found that the nonstabilizerness of the target quantum states does not affect the performance of our technique. However, the experiments are based on a limited number of target quantum circuit samples. 

Investigating the possibility of guiding PWMCTS towards solutions with low or high magic might be interesting in many domains. Depending on the goal, it can be beneficial to work with stabilizer states or to build up quantum solutions far from the classical ones.

\section{Discussion}\label{discussion}
The experiments described in previous sections demonstrate that PWMCTS successfully identifies near-optimal ansatz for variational quantum algorithms across various problems. PWMCTS shows a high level of automation given by the robustness of its results across different application domains with the fixed hyperparameter configuration. This robustness is mainly due to the reformulation of the problem and to the progressive widening technique. 
\begin{figure}[!b]
  \centering
  \includegraphics[width=1\textwidth]{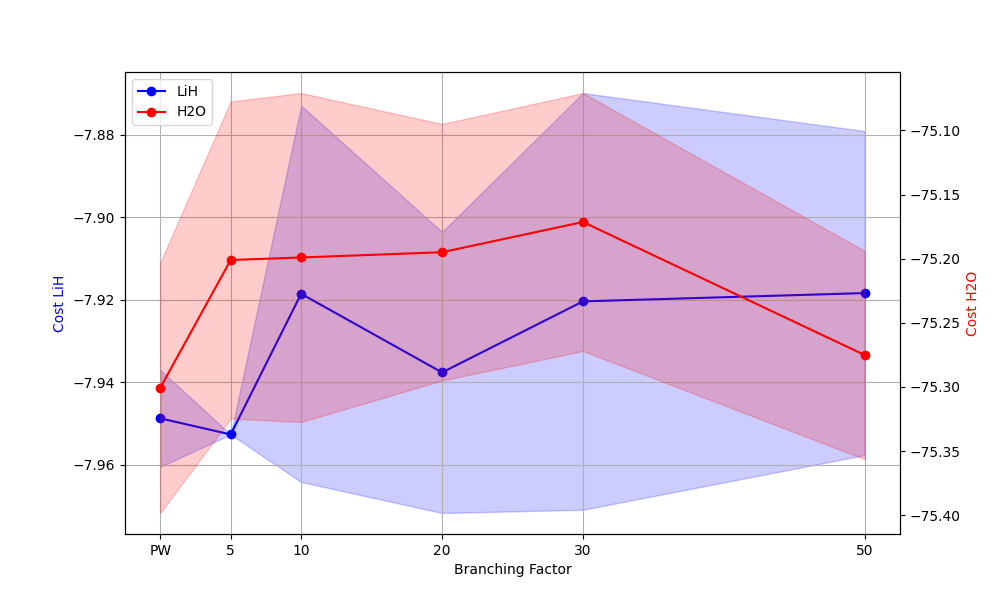}
  \caption{Analysis of branching factor. The $x$-axis represents different fixed branching factor values tested, with the first point (`PW') corresponding to experiments using the progressive widening technique. The $y$-axis shows the cost value, with two separate scales corresponding to two problems.}
\label{pw_fig}
\end{figure}
An adaptive branching factor is important as its optimal value differs substantially within the same domain. In the ground-state energy problem, the  lithium hydride molecule would require a branching factor of $5$ and the water molecule a branching factor of $50$. Figure \ref{pw_fig} shows the dependence of the best cost value on the branching factor. On the left, in blue, is the ground-state energy problem for the lithium hydride molecule. On the right, in red, is the ground-state energy problem for the water molecule. Statistics have been collected over $10$ independent runs for each branching factor value and for each problem. The scatter plot displays the mean cost values, and the shaded areas indicate the corresponding standard deviations.

The single configuration of hyperparameters, fixed for all the experiments in this article, has been determined with a grid search around reference values found in the literature \cite{franken2022quantum,sironi2019monte}. We refer the reader to Subsection \ref{mcts_qc} for further details and to Table \ref{table_hyperparameters} for the list of hyperparameters and their respective values. This hyperparameter configuration is not guaranteed to be optimal, but it is good enough to obtain relevant results in all test domains without any further problem-dependent hyperparameter search, in contrast to the previous MCTS work \cite{wang}.

Furthermore, we extended the study to an unstructured problem where the goal is to approximate a random quantum state. We experimentally demonstrate that PWMCTS approximates quantum states from a dataset of random quantum circuits. The dataset consists of quantum circuits with $4$, $6$, and $8$ qubits, containing a total number of gates ranging between $5$ and $30$, selected from the set of Clifford generator gates and the CNOT gate. It is important to note that the universal gate set used to generate the dataset of benchmark circuits is different from the one used by PWMCTS. This provides a fair benchmark, free from the underlying structural patterns between the benchmark circuits and the search technique. Additionally, we ranked the quantum states generated based on their stabilizer R\'enyi entropy to analyze the dependence of PWMCTS's performance on the nonstabilizerness of the target state. PWMCTS manages to design PQCs to approximate quantum states with different values of stabilizer R\'enyi entropy. The experiments demonstrate that PWMCTS does not exhibit any evident struggles in working with quantum states with high values of nonstabilizerness, supporting PWMCTS as a promising candidate for solving QAS problems and advancing toward quantum advantage. 

We remark that QAS is a computationally hard problem, as even the training of parameterized quantum circuits is NP-hard \cite{bittel2021training}. The search space of QAS grows exponentially with the number of qubits, and the scaling of QAS techniques is a key challenge in the field. Although the scaling of PWMCTS to larger quantum circuits and different domains of application may be a potential limitation, it has two main features that help address the scalability problem. The first lies in the balance between exploitation and exploration, originating from the intrinsic nature of MCTS, which aims to confine the search to the branches expected to be more promising rather than searching in the whole, exponentially large, tree. The second is given by the hyperparameter $d$, which restricts the search space by fixing the maximum depth of the quantum circuits explored by PWMCTS. 

Note that all the experiments, described in the previous section, are based on noiseless simulations. However, in Section \ref{noise}, we present a preliminary analysis of some experiments under the influence of quantum noise.
Section \ref{runtime_section} describes the metric used to evaluate and compare the efficiency of PWMCTS. Finally, Section \ref{comparison} compares the results obtained by PWMCTS with the state-of-the-art technique on QAS based on MCTS \cite{wang}.

\subsection{Performance Analysis under Noisy Simulations}\label{noise}
For NISQ applications, it is important to test the impact of quantum noise on the performance of PWMCTS. In this section, we extend some of the experiments of Section \ref{applications} to various simulated quantum channels. For this purpose, we worked on the ground-state energy problem for the hydrogen molecule and on the oracle approximation problem for random quantum circuits introduced in Subsection \ref{oracles}. 

PWMCTS has been tested on the ground-state energy estimation problem under bit-flip noise \cite{nielsen2010quantum}, depolarizing noise \cite{nielsen2010quantum}, and their combination. These quantum noise channels were chosen because they model noise associated with readout and gate infidelities, respectively \cite{nielsen2010quantum}. In our experiments, we fix both noise probabilities to $0.01$. Note that this value is one order of magnitude higher than the typical probabilities used to model readout errors and single-qubit gate infidelities in real quantum hardware, and is comparable to those used for two-qubit gate infidelities. The typical values for the noise probabilities are based on the IBM Ourense quantum device, which, although currently unavailable, enables comparison with previous QAS studies that investigated the influence of quantum noise \cite{du2022quantum, patel2024curriculum}.
PWMCTS has been tested on the oracle approximation problem, on circuits with four and six qubits, under different levels of bit-flip noise. Given the larger size of the problem, we did not consider depolarizing noise in this case due to computational constraints. Instead, we used significantly higher bit-flip noise probabilities: $0.1$ and $0.2$.

\begin{figure}[!t]
\centering
\begin{subfigure}{.44\textwidth}
  \centering
  \includegraphics[height=6cm]{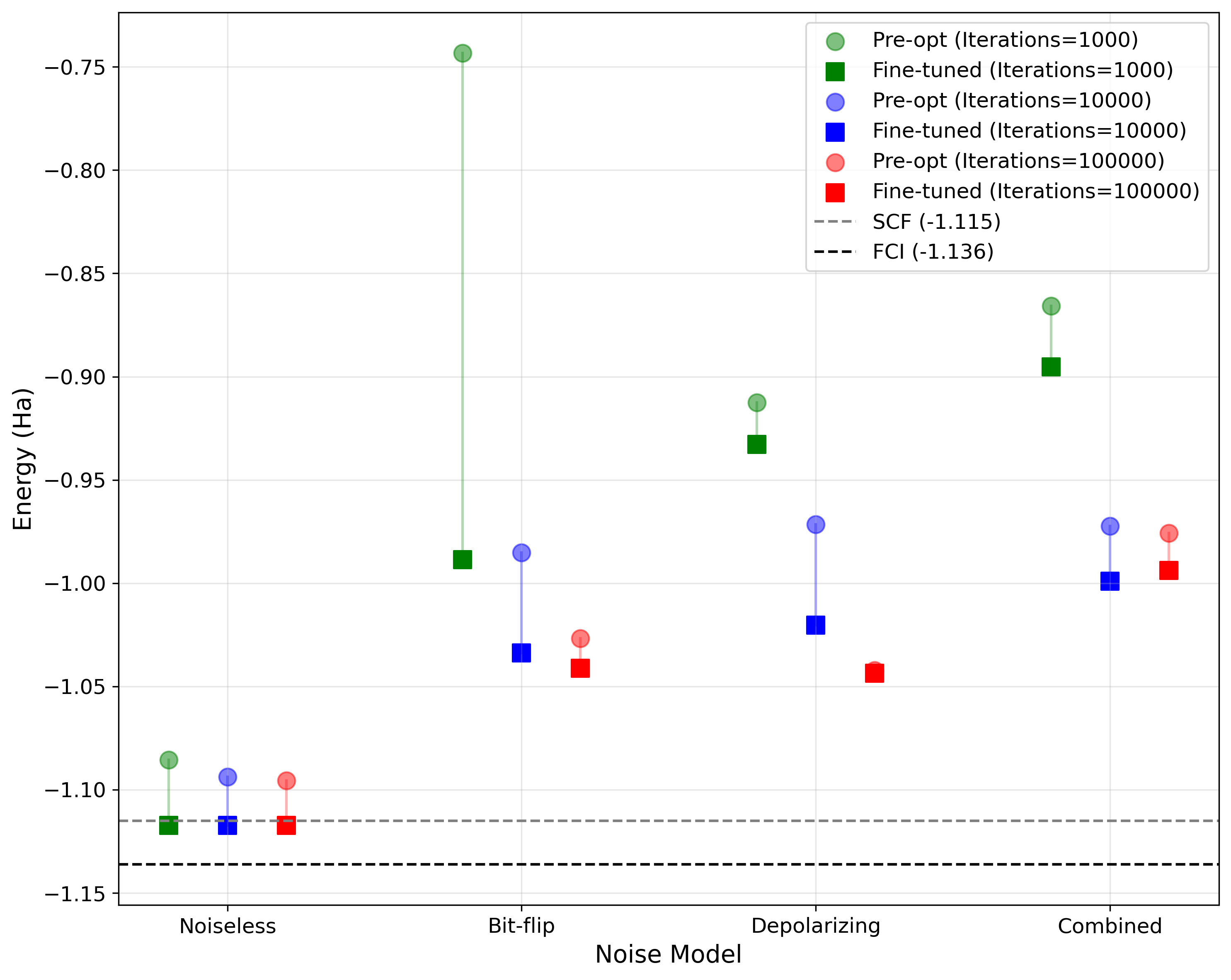}
  \caption{H$_2$}
  \label{noise_a}
\end{subfigure}
\hfill
\begin{subfigure}{.48\textwidth}
  \centering
  \includegraphics[height=6cm]{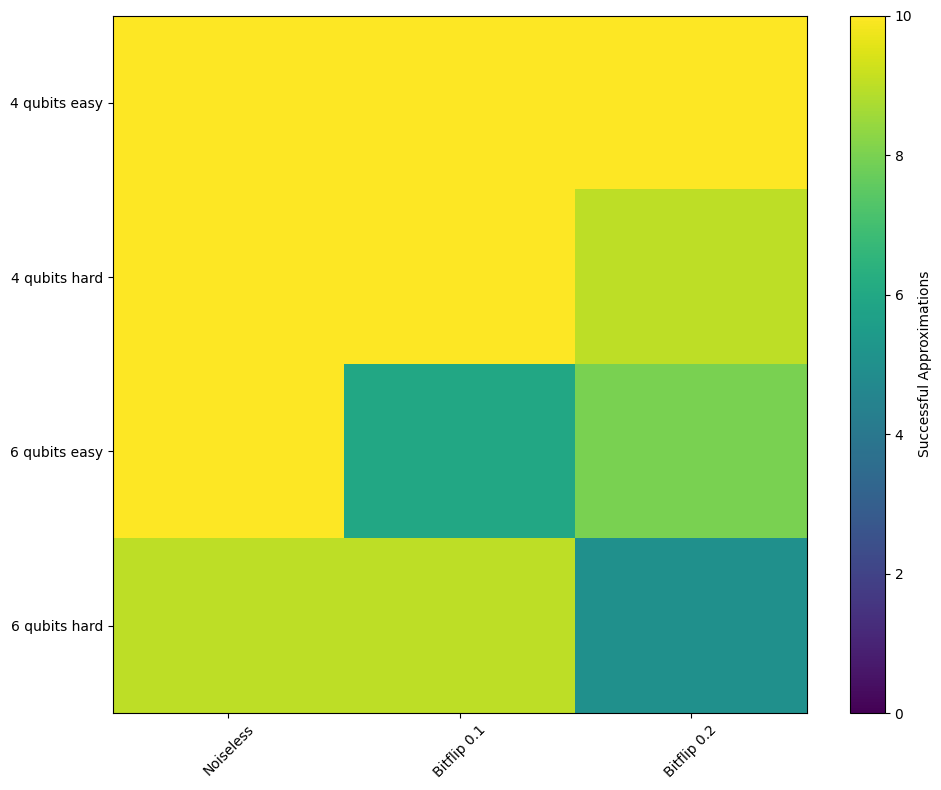}
  \caption{Random quantum circuits}
  \label{noise_b}
\end{subfigure}
\caption{Comparison of PWMCTS performance under various quantum noise channels. Figure~\ref{noise_a} presents the results for the ground-state energy estimation of the hydrogen molecule across different values of PWMCTS iterations and noise channels. Figure~\ref{noise_b} illustrates the performance on the oracle approximation problem for four random quantum circuits, defined in Subsection~\ref{oracles}, comparing different levels of bit-flip noise.}
\label{fig:noise_combined}
\end{figure}

Figure~\ref{fig:noise_combined} compares the noiseless results obtained in Section~\ref{applications} with the results under simulated noise. 
Figure \ref{noise_a}, on the left, plots the energy values for all the noise models tested and for different values of PWMCTS iterations $I$ ($I=1000$ in green, $I=10000$ in blue, $I=100000$ in red). The circles indicate the results of the best node (quantum circuit) in the tree, while the squares represent the results after the fine-tuning of the parameters with the Adam optimizer.
Figure \ref{noise_a} shows that PWMCTS does not manage to compete with the results of the classical SCF and FCI benchmarks. However, increasing the number of PWMCTS simulations significantly improves the results. As expected, the worst results are obtained on the noise model that combines bit-flip and depolarizing noise. On the one hand, the results obtained by PWMCTS with $I=10{,}000$ and $I=100{,}000$ improve upon those reported in \cite{du2022quantum}. On the other hand, the reinforcement learning approach proposed for QAS in \cite{patel2024curriculum} manages to reach the same energy as FCI within chemical accuracy. However, this latter approach has only been tested in the quantum chemistry domain and incorporates domain knowledge through machine learning. In contrast, our approach is completely domain-agnostic. We note that, for a given problem, the number of gates and parameters required in the quantum circuit to solve it is expected to increase with the noise level \cite{sharma2020noise,fontana2021evaluating,patel2024curriculum}. Therefore, in our case, to find the optimal circuit in the presence of noise, we should increase the value of the hyperparameter $d$, which constrains the depth of the circuit.

Figure \ref{noise_b}, on the right, shows the influence of bit-flip noise on the performance of PWMCTS in approximating four of the quantum circuits defined in Section \ref{oracles}. Specifically, we considered the four circuits with the number of gates fixed to $20$, denoted as ($n=\{4,6\}$, $g=20$, $M_2=\{\text{easy},\text{hard}\}$).  Performance is measured in terms of the number of successful approximations according to the definition given in Section \ref{oracle_experiments}, with the same $\epsilon$ values.
The first column in Figure~\ref{noise_b} reports the same results shown in Figure~\ref{oracle_figure}, while the second and third columns refer to  experiments with increasing bit-flip noise probability. The first two rows indicate that for the easiest problems, $4$-qubit quantum circuits, PWMCTS is significantly robust. It only fails to approximate the circuit in 1 out of 10 independent run ($4$, $20$, $\text{hard}$) under the highest noise level. Similarly, the last row, related to the 6-qubit circuit with the highest nonstabilizerness, shows that PWMCTS obtains the same number of successful approximation as in the noiseless case and it only decreases of one for the highest noise level.

QAS in the noiseless scenario can be considered a single-player game in deterministic environments. In the presence of noise multiple evaluations of the cost function on the same circuit may bring different values.
QAS in the noisy scenario can be considered a single-player game in a stochastic environment. We remark that the results obtained in this section are based on exactly the same implementation of PWMCTS, including the hyperparameter configuration. However, PWMCTS would benefit from MCTS refinements to better cope with this change. Several MCTS variants have been proposed to handle stochastic single-player games \cite{browne2012survey, swiechowski} and we believe it is worthy to further investigate in this direction.

We emphasize that PWMCTS requires further studies before it can be reliably used to design quantum circuits for real quantum hardware. Nevertheless, two main reasons motivate the consideration of PWMCTS as a promising candidate for tackling QAS on real hardware. First, there is a need of gradient-free techniques, which are usually more robust to noise-induced barren plateaus \cite{wang2021noise}. Second, extensive research has already been conducted in the classic domain of games, where MCTS behaves well even in stochastic settings \cite{browne2012survey, swiechowski}. The results presented in this article are promising in this regard, especially considering that we did not incorporate any enhancements to MCTS for handling stochastic environments.

\subsection{Runtime Analysis}\label{runtime_section}
The performance analysis of QAS techniques takes into account the value of the cost function on the quantum circuit designed, the count of CNOT gates and parameterized gates, the robustness of the hyperparameter configurations, and runtime.
The value of the cost function for the designed PQC measures the quality of the solution to the original problem compared to the optimal cost value. The count of CNOT and parameterized gates reflects the efficiency and complexity of the ansatz. A lower count of CNOT gates is desirable to execute the circuit on NISQ devices reliably, as the error rate of single-qubit gates is typically negligible compared to that of two-qubit gates. The count of parameterized gates affects the optimization difficulty of the ansatz. The robustness of the hyperparameter configuration of QAS techniques measures the degree of automation and adaptability to different problems and domains of application.

The runtime analysis is performed in terms of the number of total quantum circuit evaluations $N_{eval}$. It refers to the number of quantum circuits executed on a quantum device to evaluate the cost function until a circuit providing a sufficiently good solution is successfully designed. The number of quantum circuit evaluations required by PWMCTS is 
\begin{equation}
    N_{eval} = I + 2 \times l \times T .\label{runtime}
\end{equation}
It consists of two terms. The first relates to the gradient-free step performed with the tree search, while the second relates to the final fine-tuning with the Adam optimizer. We recall that $I$ is the number of PWMCTS iterations, $l$ is the number of parameterized quantum gates in the PQC, and $T$ is the number of optimizer steps used to fine-tune these $l$ circuit parameters. 

According to previous work on QAS \cite{wang}, the runtime analysis is based on the number of quantum circuit evaluations. This serves as the reference comparison metric, as it is independent of hardware, software libraries, and the specific quantum circuits explored during the search. 
We note that the progressive widening technique \cite{couetoux2011continuous} does not result
in significant computational overhead for MCTS.

\subsection{Comparison with Nested Monte Carlo Tree Search}\label{comparison}
We benchmark PWMCTS with the Nested Monte Carlo Tree Search (NMCTS) approach proposed in \cite{wang} for two main reasons. First, both techniques inherently represent the QAS problem as a tree and solve it using an MCTS approach with an identical reward function. Second, NMCTS has been extensively tested in various domains and with up to $10$ qubits, making it an appropriate testbed. In contrast, the related works introduced in Section \ref{related_works} based on different techniques have been mainly tested on small-scale problems within a limited set of application domains. 
\begin{table}[!b]
    \centering
    \caption{Comparison of the performance of PWMCTS and NMCTS \cite{wang} on quantum chemistry applications (H$_2$, H$_2$O, LiH), and on a system of linear equations (VQLS). We compare the best quantum circuits found by both techniques in terms of result, which is the cost value, the number of quantum circuit evaluations required to design the quantum ansatz, $N_{eval}$, and their CNOT and angle parameter counts.}
    \label{table_comparison}
    \begin{tabular}{
        @{} >{\centering\arraybackslash}m{2.8cm}  
        >{\centering\arraybackslash}m{2cm}      
        >{\centering\arraybackslash}m{2cm}      
        >{\centering\arraybackslash}m{2cm}      
        >{\centering\arraybackslash}m{2.8cm}      
        @{}
    }
        \toprule
         & H$_2$  & H$_2$O & LiH & VQLS \\
        \midrule
        Qubits $n$ & 4  & 8  & 10  & 4  \\
        \midrule
        \multicolumn{5}{c}{\textbf{PWMCTS}} \\
        \midrule
        Result & $-1.117$  & $-75.422$  & $-7.9526$  & $3.98\times 10^{-8}$  \\
         $N_{eval}$ & \textbf{4200}  & \textbf{14500}  & \textbf{12360}  & \textbf{10780}  \\
        \# CNOTs   & \textbf{2}  & \textbf{6}  & \textbf{1}  & 4  \\
        \# Parameters     & \textbf{10}  & \textbf{19}  & \textbf{20}  & \textbf{6}  \\
        \midrule

        \multicolumn{5}{c}{\textbf{NMCTS \cite{wang}}} \\
        \midrule
        Result & $-1.117$  & $-75.422$  & $-7.9526$  & $9.85 \times 10^{-6}$  \\
         $N_{eval}$  & 1337500  & 1095000  & 297000  & 263250  \\
        \# CNOTs   & 13  & 18  & 5  & \textbf{3}  \\
        \# Parameters $\theta$    & 27  & 60  & 24  & 21  \\
        
        \bottomrule
    \end{tabular}
\end{table}
Although both NMCTS and PWMCTS address QAS using MCTS, they differ fundamentally in the tree representation of the problem and the MCTS variant employed. NMCTS structures the tree representation by slicing the target quantum circuit into layers. The layers are filled in by the Nested MCTS. The search of the parameters is not considered at all by the NMCTS but is carried out by a classical Adam optimizer. It is performed using a weight-sharing technique and estimating average gradients. In contrast, PWMCTS structures the tree to search for both the structure (topology) and the parameters of the circuit simultaneously. In this formulation, the tree depth is not directly related to the circuit depth. 
Moreover, the universal gate sets used by the two approaches are different. PWMCTS works with four gates: the CNOT gate and the three single-qubit rotation gates \( R_x \), \( R_y \), and \( R_z \), applied independently.  
In contrast, NMCTS works with two gates: the CNOT gate and a single-qubit rotation gate \( Rot(\theta_x, \theta_y, \theta_z) = R_x(\theta_x) R_y(\theta_y) R_z(\theta_z) \), which rotates the qubit simultaneously along all three axes.

\noindent Table \ref{table_comparison} summarizes the experimental results of the two techniques on four problems. 
The runtime analysis highlights a significant improvement in the number of circuit evaluations required by PWMCTS, achieving a reduction of $10$ to $100$ times compared to NMCTS while delivering the same or better results. The PQCs designed by PWMCTS are also shallower in terms of both CNOT and angle parameter count, up to a factor 6 and 3, respectively. Moreover, PWMCTS achieves these results using a single hyperparameter configuration, while NMCTS requires tuning several hyperparameters, for which the values are different for each application.

\section{Conclusions}\label{conclusions}
This article investigated the role of MCTS in designing PQCs for the quantum architecture search problem. The main goal was the development of a problem-agnostic MCTS technique that could reduce the computational cost compared to previous MCTS research \cite{wang}. For this purpose, we have proposed PWMCTS, which is a problem-agnostic, hardware-tailored, and gradient-free technique for QAS. It consists of two contributions. The first is a different formulation of the QAS problem in the tree, where each tree node represents a quantum circuit and each edge represents an action that modifies the quantum circuit. These actions are defined by sampling in a set of PQC modifications, which consists of variation in the circuit structure or of a parameter value \cite{franken2022quantum}. The second contribution equips MCTS with a progressive widening technique \cite{couetoux2011continuous}. It is employed to deal with an infinite number of possible actions, as it adapts the branching factor along the tree. Progressive widening provides MCTS with a higher level of automation reducing the number of hyperparameters to tune across different problems. 

Experimental results show that PWMCTS converges to near-optimal ansatz in several application domains, including quantum chemistry, systems of linear equations, and quantum oracle approximation. It provides improvements in terms of both automation and computational resources. PWMCTS does not need prior knowledge of the problem or preliminary hyperparameter tuning. Moreover, the technique requires a considerably lower number of circuit evaluations and shallower quantum circuit designs in terms of both the number of CNOTs and angle parameters. 

\noindent The generality of this approach comes not only from the few hyperparameters to be tuned, but also from its adaptability to different problems. In principle, it can be employed not only to build up quantum circuits but also for quantum circuit optimization tasks \cite{karuppasamy2025comprehensive}. In fact, by defining an appropriate reward function, we could start the search from the quantum circuit to optimize (the root of the tree) and set the probability distribution that defines the action space such that PWMCTS only searches for shallower circuits.

We identify four main directions for future research. 
The first deals with the robustness PWMCTS under the influence of quantum noise. The preliminary experiments show promising results for the performance of PWMCTS under various noise models, both for the ground-state energy estimation of the hydrogen molecule and for the oracle approximation problem. However, it is an interesting direction to include MCTS enhancements developed for stochastic settings \cite{browne2012survey,swiechowski}, to QAS. Moreover, we also highlight the implementation of PWMCTS on real quantum hardware as a future direction.
The second future direction is extending PWMCTS to other domains, such as combinatorial optimization and machine learning. For the former, MCTS can be used to design ansatz to solve NP-hard problems, like MaxCut, with a VQE approach. Moreover, it can be used as a gradient-free technique to optimize the parameters of the PQCs to those problems with a Quantum Approximate Optimization Algorithm \cite{agirre2024monte}. For the latter, PWMCTS might be used to design quantum kernels and also to address overfitting issues in quantum neural networks \cite{peters2023generalization,scala2023general}.
The third future direction deals with the design and integration of problem-dependent heuristics or machine learning models in the roll-out and expansion steps \cite{winands2019monte}. In principle, PWMCTS may be applied to any QAS problem. However, as stated by the free-lunch theorem, it cannot compete with independent problem-tailored approaches. It is an open question whether our approach can achieve state-of-the-art results in specific domains by providing heuristics or training machine learning models to guide the search towards more promising states. The fourth future direction deals with the MCTS enhancements \cite{swiechowski}. Although MCTS enhancements have been extensively studied in the domain of games, their role in the domain of quantum computing is still not well investigated. Moreover, quantum enhancements for MCTS are worthy of investigation \cite{lumbreras2022multi}.


\bigskip
\noindent \textbf{Data Availability}\\
The code and supplementary experiments are accessible via Github 
and the source code archived in Zenodo at time of publication. 
\begin{itemize}
    \item GitHub: \href{https://github.com/VincenzoLipardi/MCTS-QAS}{https://github.com/VincenzoLipardi/MCTS-QAS}
    \item 
Zenodo: \href{https://doi.org/10.5281/zenodo.14802441}{https://doi.org/10.5281/zenodo.14802441}
\end{itemize}

\medskip
\noindent 
\textbf{Acknowledgements} \\
The authors acknowledge the anonymous reviewers for their helpful comments and suggestions that improved the manuscript, especially regarding the experiments with simulated noise.

\bigskip
\noindent \textbf{Conflict of Interest}\\
The authors have no competing interests to declare that are relevant to the content of this article.
\medskip

%
\bibliographystyle{unsrt}

\bibliography{bibliography}




\end{document}